\documentclass[prb,superscriptaddress,aps, bibliography, twocolumn, reprint, showpacs, footnoteinbib]{revtex4-1}

\usepackage{color}    
\usepackage{transparent}   
\usepackage{units}  
\usepackage{float} 
\usepackage{wrapfig}

\usepackage{graphicx}
\usepackage{amssymb}   
\usepackage{amsfonts}
\usepackage{amsmath}
\usepackage{dsfont}
\usepackage{natbib}
\usepackage{dcolumn}   
\usepackage{bm}        
\usepackage[mathscr]{eucal}
\usepackage[dvipsnames]{xcolor}
\usepackage[colorlinks, linkcolor=OrangeRed,citecolor=RoyalBlue,urlcolor=NavyBlue]{hyperref}

\usepackage[all]{hypcap} 
\usepackage{mathtools}
\usepackage{dsfont}

\definecolor{myblue}{rgb}{0,0,0.75}

\usepackage{graphicx}
\usepackage{amsmath,amssymb,bm}
\usepackage{enumerate}
\usepackage{braket}        

\begin{document}

\title{Distinguishing an Anderson Insulator from a Many-Body Localized phase through space-time snapshots with Neural Networks}
\author{Florian Kotthoff}
\affiliation{Department of Physics, Technische Universit\"at M\"unchen, 85747 Garching, Germany}
\author{Frank Pollmann}
\affiliation{Department of Physics, Technische Universit\"at M\"unchen, 85747 Garching, Germany}
\affiliation{Munich Center for Quantum Science and Technology (MCQST), Schellingstr. 4, D-80799 M\"unchen, Germany}
\author{Giuseppe De Tomasi}
\affiliation{Cavendish Laboratory, University of Cambridge, Cambridge CB3 0HE, United Kingdom}

\begin{abstract}
Distinguishing the dynamics of an Anderson insulator from a Many-Body Localized (MBL) phase is an experimentally challenging task. In this work, we propose a method based on machine learning techniques to analyze experimental snapshot data to separate the two phases. We show how to train $3D$ convolutional neural networks (CNNs) using space-time Fock-state snapshots, allowing us to obtain dynamic information about the system. We benchmark our method on a paradigmatic model showing MBL ($t-V$ model with quenched disorder), where we obtain a classification accuracy of $\approx 80 \%$ between an Anderson insulator and an MBL phase. We underline the importance of providing temporal information to the CNNs and we show that CNNs learn the crucial difference between an Anderson localized and an MBL phase, namely the difference in the propagation of quantum correlations. Particularly, we show that the misclassified MBL samples are characterized by an unusually slow propagation of quantum correlations, and thus the CNNs label them wrongly as Anderson localized.
Finally, we apply our method to the case with quasi-periodic potential, known as the Aubry-Andr\'e model (AA model). We find that the CNNs have more difficulties in separating the two phases. We show that these difficulties are due to the fact that the MBL phase of the AA model is characterized by a slower information propagation for numerically accessible system sizes.  
\end{abstract}
\maketitle
\section{Introduction}

Advancements of controlled experimental techniques, such as ultra-cold atoms in optical lattices, trapped ions, and superconducting q-bits have led to considerable interest in the out-of-equilibrium dynamics of isolated quantum many-body systems~\cite{Gross995,Monroe_2021,Blatt2012,Kjaergaard_2020,Bloch_review_2008}. In particular, it became possible to provide experimental evidence of many-body localization (MBL)~\cite{Schreiber2015Coldatoms,Smith2016QSimulator,Choi2016Coldatoms}, and therefore to shed light upon the emergence of the laws of statistical mechanics in the quantum realm. 

\begin{figure}[h!]
\fontsize{8pt}{10pt}\selectfont
\centering
\def\svgwidth{.45\textwidth}
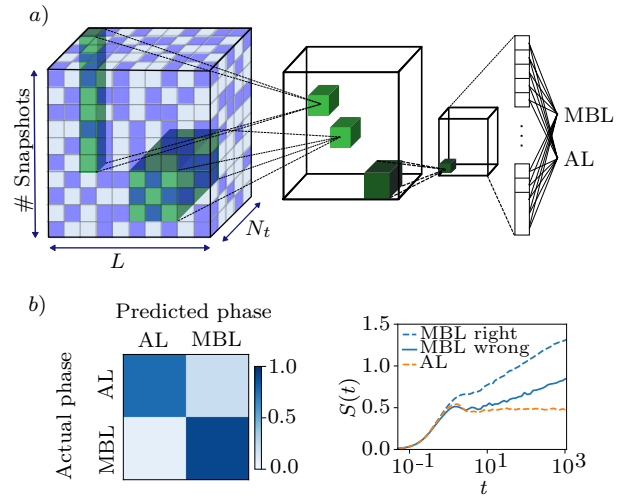
\caption{(a) Three dimensional ($3D$) structure of the input data ($\# \text{Snapshots} \times L \times N_t$) with a pictorial representation of the CNN used to distinguish an MBL phase from an Anderson localized one. (b) The left panel shows the confusion matrix for the classification task. The CNN classifies $83.9 \%$ of AL and MBL samples correctly. The right panel shows the entanglement entropy $S(t)$ averaged over MBL samples ($V=1$), right and wrong classified separately, and $S(t)$ for the non-interacting case ($V=0$). As one can notice for the wrongly classified MBL samples, $S(t)$ has an unusual slower growth of the entanglement.}
\label{fig:InceptionNN}
\end{figure}

MBL generalizes Anderson localization to the interacting case and has emerged as a novel paradigm for ergodicity breaking of generic many-body systems subjected to strong disorder~\cite{Huse_review_2015, Altman_review_2015, Abanin_RevModPhys_2019, ALET_review_2017, Imbrie_2017}. 
An MBL phase is best understood in terms of an emergent form of integrability, meaning that the system is fully described by an extensive number of quasi-local integrals of motion (LIOMs), which are adiabatically connected to the non-interacting ones. As a result, due to the local nature of the LIOMs, transport in the system is strongly hindered~\cite{DeTomasi2019Efficient,Huse_2014_fully,Local_Serbyn_2013,Chandran_2015}. Furthermore, unlike an Anderson insulator, 
interactions weakly couple the LIOMs, producing a dephasing mechanism that yields a slow logarithmic spread of entanglement~\cite{Bardarson2012unbounded, Serbyn2013universal, Prosen_2008}. Among the propagation of entanglement entropy, other dynamical indicators have been proposed to distinguish an Anderson insulator from an MBL phase, i.e., propagation of quantum mutual information~\cite{DeTomasi2017QMI}, quantum Fisher information~\cite{Smith2016QSimulator,DeTomasi2019Efficient}, temporal fluctuations~\cite{DeTomasi2019Efficient, Serbyn2014quenches} or spin noise spectroscopy~\cite{Rajeev_2015, Serbyn_inter_2014}. However, it remains a challenge to reliably distinguish the two phases experimentally. These difficulties come mainly from the fact that the distinction requires the measurement of non-local quantum correlations.

The aim of this work is to show how to use machine learning toolboxes and, in particular, convolutional neural networks (CNNs) to analyze experimental snapshot data and distinguish an MBL phase from an Anderson insulator.

Machine learning techniques have proven to be a useful tool in characterizing and understanding correlations in quantum phases of matter~\cite{Nieuwenburg2017Confusion,Nieuwenburg2019Bloch,Carrasquilla2017MLPhase,Wetzel2017Unsupervised,Chng2017Correlated, Carleo2017Solving,Zhang2018Topology,Bohrdt2019Snapshots, Robert_2020,Titus_2020}.
Recently, several works used machine learning to investigate the MBL transition which separates a thermal phase from the localized one~\cite{Doggen2018LargeChains,Hsu2018Elusive,Huembeli2019Automated,Rao2018Random,Schindler2017Probing,Kausar2020Learning,Theveniaut2019Precise,Zhang2019Interpretable, Nieuwenburg2018PTDynamics}. In particular, in Ref.~\onlinecite{Bohrdt_2020} a CNN was trained using experimentally Fock-space snapshots data to distinguish an MBL from a thermal phase. However, the distinction of an Anderson insular from an MBL phase is more subtle than the separation of an ergodic/extended from an MBL phase. 

In both, the Anderson insulator and MBL phase, degrees of freedom are frozen and the only difference is found in the propagation of quantum correlations, which are harder to be measured. By using \textit{space-time} Fock-space snapshots as input data, we show that CNNs are able to capture the important spatial and temporal correlations to distinguish the two phases. This approach should be opposed to the ones used to separate an ergodic from an MBL phase~\cite{Bohrdt_2020}, in which only Fock-space snapshots at a single time are sufficient to distinguish the two phases. 

Concretely, in quantum simulation platforms such as ultra-cold atoms in optical lattices, Fock-space snapshots are accessible following a time evolution with the use of a quantum microscope~\cite{Bakr2009, Sherson2010}. The computation of local observables at a specific target time is then found by a proper average over the ensemble of snapshots. Here, we show how to construct and train a CNN (see Fig.~\ref{fig:InceptionNN}) to extract dynamic properties and therefore to distinguish the two phases. Importantly, we provide numerical evidence that only a reasonable amount of snapshots is needed to separate the two phases, which bounds the number of experimental measurements. Moreover, we show that the MBL samples that are wrongly classified are characterized by an atypical slow propagation of information, i.e., entanglement entropy and particle number fluctuation, see Fig.~\ref{fig:InceptionNN}$b$. This supports the idea that the CNN learns the important aforementioned features to distinguish an MBL phase from an Anderson localized one.

This work is structured as follows. In Sec.~\ref{sec:Model}, we introduce the utilized models. In Sec.~\ref{sec:Method}, we explain the structure of the neural network and the generation of the snapshots. Section~\ref{sec:NetworkPerformance} is dedicated to examining the network performance on classifying an Anderson insulator and an MBL phase. 
In particular, we show that it is necessary to consider snapshots from different points in time and thus gain dynamic information to improve the classification accuracy. 
In Sec.~\ref{sec:robustness}, we test the stability of our neural network by tuning the chain length and interaction strength of the input data. The observed stability of the trained neural network is an important component of our work. It provides an indication that our method can be applied to real experimental data. With the aim to understand the high performance of our network, in Sec.~\ref{subsec:ComprehendHubbard} we show that the wrongly classified MBL samples are characterized by an unusually slow growth of entanglement, see Fig.~\ref{fig:InceptionNN}$b$. Thus, our CNNs are able to detect the important features that distinguish an Anderson insulator from an MBL phase. Finally, we apply our method to the case in which the disorder is generated by a quasi-periodic potential (Aubry-Andr\'e model), which is particularly relevant for experiments~\cite{Fangzhao_2020,Schreiber2015Coldatoms, Slow_Altman_2017, Bordia_2017,Bordia_2016, Kohlert_2019}. 

\section{Model} \label{sec:Model}
We study the $t-V$ disordered spinless fermionic chain with periodic boundary conditions
\begin{align}
\label{eq:Hamiltonian}
\begin{split}
H = & -\frac{t}{2}\: \sum_{j=1}^{L} \: c^\dagger_jc_{j+1} \: + h.c. \: + \: \sum_{j=1}^L h_j \: \left(n_j - \frac{1}{2} \right) \\ 
& + \: V \: \sum_{j=1}^{L} \: \left( n_j - \frac{1}{2} \right)\left( n_{j+1} - \frac{1}{2} \right),
\end{split}
\end{align} 
 where $c^\dagger_j$ $(c_j)$ is the fermionic creation (annihilation) operator at site $j$. $t=1$ and $V$ are the hopping and the interaction strength, respectively, and $\{h_i\}$ are  random fields which are uniformly distributed between $[-W,W]$. $L$ is the length of the chain and $N = L/2$ is the number of fermions (half filling). 
 
 For $V=0$, all the single-particle wavefunctions are exponentially localized for any amount of disorder~\cite{Anderson1958Loc,Evers2008Review,Mott1961theory}. The interacting case ($V\ne 0$) is the paradigmatic model which is believed to have an MBL transition. Several numerical works have shown that the critical value of the transition is $W_c \approx 3.5$ for $V=1$~\cite{Luitz15Edge,Tomasi17Mutual,Bera15OnePart,Serbyn2015criterion,Serbyn2014quenches, Pal_2010,Oga_2007} ($W<W_c$ ergodic and $W>W_c$ localized).
 
Additionally, in the final part of our work, we will test our method on the quasi-periodic case, known as the Aubry-Andr\'e model (AA model)~\cite{Aubrey_1980}. The AA model is obtained from Eq.~\ref{eq:Hamiltonian} by setting $t=1$, $h_j= W \cos(2\pi j\phi + \alpha)$, where $\phi = \frac{1+\sqrt{5}}{2}$ is the golden ratio and $\alpha$ a random phase uniformly distributed between $[0,2\pi]$. For $V=0$, the AA model has a metal insulator transition at $W_c=1$~\cite{Aubrey_1980}, extended for $W<W_c$ and localized for $W>W_c$. For $V=1$ the AA model is believed to show MBL at strong disorder ($W_c>4$)~\cite{Shankar_2012}.

In order to be in the strongly localized regime, in both models, we consider $W\in \{6,7,8\}$. 

\section{Method}
\label{sec:Method}
In this section, we introduce the numerical methods and the structure of the CNN \cite{rawat2017deep} (see Fig.~\ref{fig:InceptionNN}). As shown schematically in Fig.~\ref{fig:InceptionNN}~(a), the first layer is an adapted version of an inception layer~\cite{Szegedy2014Inception}, followed by a convolution layer and two fully connected layers. As we will show, this architecture enables the network to achieve good results in classifying the snapshots. A more detailed description of the architecture and the hyperparameters can be found in the Appendix.

\begin{figure}[h!]
\includegraphics[width=.95\columnwidth ,trim={.7cm 1cm 1cm 1cm},clip]{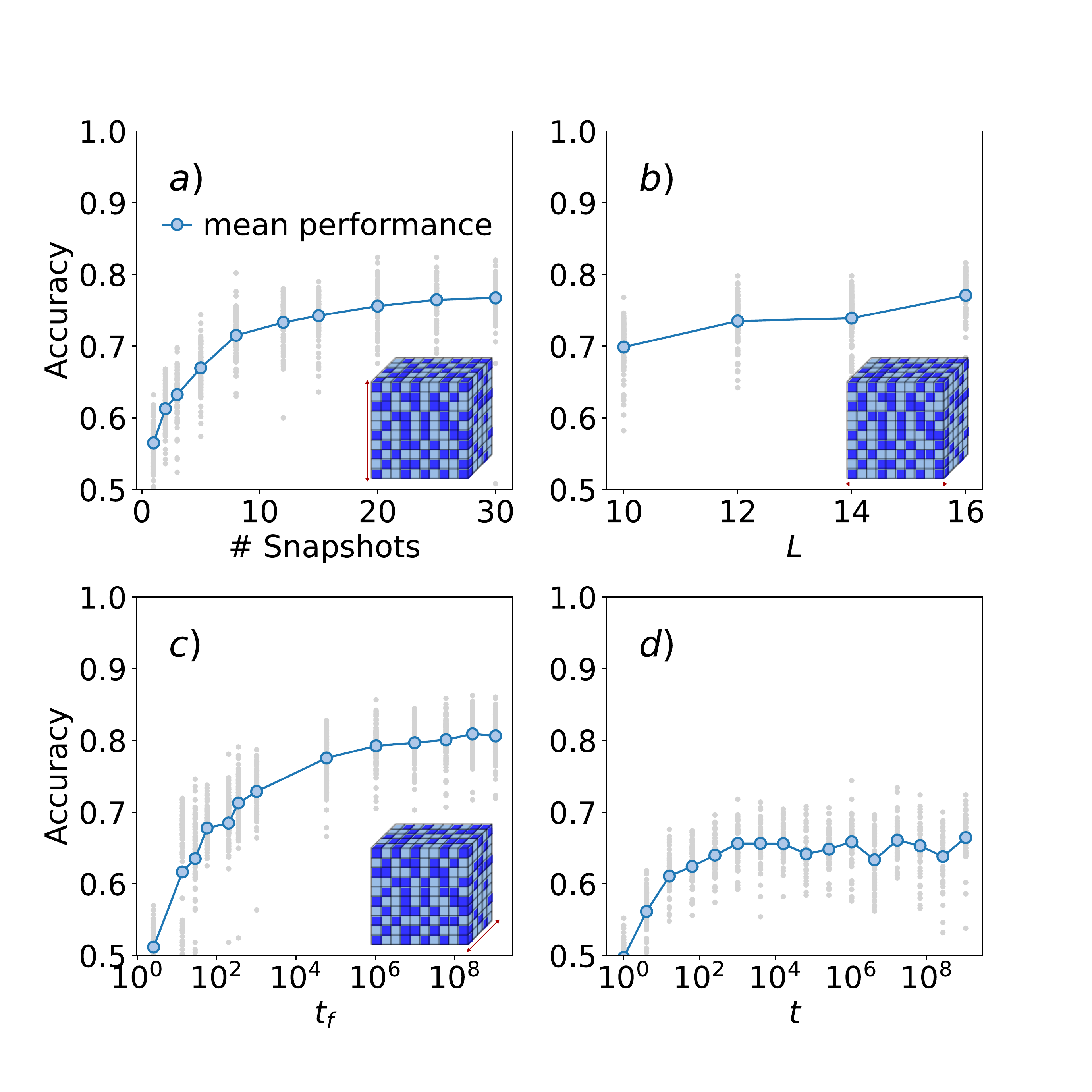}
\caption{Dependence of the classification accuracy on variations of the snapshot block parameters $\#\text{Snapshots}$, $L$, $t_f$ and $t$, where we tune each parameter separately. (a): $L =16$, $t_f = 10^3$ and we tune $\#\text{Snapshots}$. (b): $\#\text{Snapshots}=30$, $t_f = 10^3$ and $L$ is varied. (c): $\#\text{Snapshots}=30$, $L = 16$ are fixed and the final time $t_f$ is varied. (d) Unlike the other panels where the CNNs were trained using snapshots at different times in the range $[0,t_f]$, in this panel, the CNNs were trained using only snapshots at a single target time $t$. In all panels, the grey dots are the performance of single CNNs and $N_t=11$ is fixed.}
\label{fig:XXZ_BlockDependence}
\end{figure}

The Fock-state snapshots have been obtained from the out-of-equilibrium dynamics of the system using exact diagonalization.  In particular, with the aim to get as close as possible to an experimental set up, we follow the dynamics of a global quench starting from the N\'eel state $| \psi \rangle = \prod_{j}^N c_{2j}^\dagger |0\rangle$. During the dynamics, we compute the probability to be in the Fock-state $|\underline{n} \rangle = \prod_{i}^L (c_i^\dagger)^{n_i} |0\rangle $ with $n_i \in \{0,1\}$
\begin{equation}
\label{eq:Probability}
P_{\underline{n}}(t) = |\langle \underline{n} |  e^{-it H}|\psi  \rangle |^2.
\end{equation}
Finally, at each target time $t$, a given number of Fock-state snapshots $\underline{n} = (n_1, n_2, \cdots, n_L)$ are sampled from $P_{\underline{n}}(t)$. Doing so, we simulate the collapse of the wavefunction, as done in an experimental protocol using a quantum microscope~\cite{Bakr2009, Sherson2010}. 

We organise the input data for the neural network in a three-dimensional ($3D$) structure as depicted in \mbox{Fig.~\ref{fig:InceptionNN}~(a):} $\# \text{Snapshots} \times L \times N_t$. A chosen number of Fock-state snapshots are taken at fixed target times and then stacked together to form two-dimensional slices. Along the $t$-axis, slices from different times are then concatenated to form the $3D$ input blocks for  CNNs (see \mbox{Fig.~\ref{fig:InceptionNN}~(a)).} The resulting $3D$ block structure contains information that can be measured in experiments, namely the location of the fermions at a given time. If not stated otherwise, we will consider $\# \text{Snapshots} \times L \times N_t = 30 \times 16 \times 11$, where $30$ snapshots were taken at $11$ different times from systems of size $L=16$. The target points in time are distributed linearly in log scale up to a final time $t_{f}$.

Finally, we repeat the procedure above to obtain one snapshot block for each random instance of the fields $\{h_j\}$. Then, we merge all snapshot blocks and randomly divide them into training, validation, and test set, as it is usually done in classification tasks (cross-validation)~\cite{Goodfellow2016DLBook}. The network is trained using only data from the training set and adjusts its weights depending on the snapshots it uses in the training. 
To avoid over-fitting on the training set, the network classifies the validation set after each training epoch. The CNN that achieves the best accuracy on the validation set is used after training to classify the test set~\cite{Goodfellow2016DLBook}. This procedure is repeated for $100$ CNNs (if not stated otherwise) to get an averaged classification accuracy.


\section{Neural network performance} \label{sec:NetworkPerformance}
In this section, we study the performance of our neural network in distinguishing the MBL phase from the Anderson localized one. Furthermore, we inspect the dependence on the input data parameters, i.e., $\# \text{Snapshots}$, system size $L$ and final time $t_f$. Particular attention is given to the $3D$ structure of our data and on the fundamental importance of the time axis, which provides dynamic information of the system. 

The performance of our CNN is quantified by the accuracy $\in [0,1]$, which is given by the number of exactly classified samples within the test set divided by the total number of samples in the test set. Thus, the CNN fails to classify the two phases if the accuracy is $\approx 1/2$ since it is equivalent to making a decision using a fair coin. Otherwise, if the accuracy is equal to one, the neural network has a perfect performance. Classifying ergodic/thermal and MBL phases with neural networks can be done with high accuracies $\approx 1$\cite{Bohrdt_2020,Theveniaut2019Precise,Zhang2019Interpretable,Hsu2018Elusive} since the two phases are fundamentally distinct. However, one expects that distinguishing an MBL from the Anderson  phase is a much harder problem, since the 
two phases are both localized and differ only in terms of the information propagation~\cite{Serbyn_2014_local, Bardarson2012unbounded}. 

First, we study the dependence of our classification accuracy on the value of $\# \text{Snapshots}$ used in the CNN. This is particularly relevant in an experimental setup, where one would like to reduce the number of measurements. 
Figure~\ref{fig:XXZ_BlockDependence}$a$ shows the mean test accuracy for $t_f=10^3$ and $L=16$ as function of $\# \text{Snapshots}$, where the average was performed over $100$ CNNs. For completeness, in Fig.~\ref{fig:XXZ_BlockDependence}, the performance of individual CNNs is also reported (dashed vertical dots), which quantifies the error. 
As expected, the accuracy increases with the number of snapshots used and reaches its maximum at $\approx  0.75$. Importantly, the accuracy reaches an almost constant value already for $\# \text{Snapshots}=15$, meaning that only a limited number of measurements are needed. In Fig.~\ref{fig:XXZ_BlockDependence}~(b), the accuracy of our CNNs is shown as a function of the system size $L$. Although only slightly, the mean accuracy increases with $L$, providing evidence that in an experimental setup, where hundreds of sites can be probed, the CNNs could perform even better. 

Another important parameter of our input data is given by the final time $t_f$, which is the latest time in the dynamics. In Fig.~\ref{fig:XXZ_BlockDependence}~(c), we study the dependence of $t_f$ on our results. In particular, we train our CNNs always taking $N_t = 11$ target times in $t \in (0, t_f]$. \footnote{The target times are log-linear distributed.} This ensures that the amount of input data is independent of the final time $t_f$, and consequently, we can fairly compare the performance of our CNNs at different $t_f$. An increase in accuracy as a function of $t_f$ is observed (see Fig.~\ref{fig:XXZ_BlockDependence}~(c)), reaching $\text{accuracies} \approx 0.8$ for $t_f \approx 10^8$. Though these large times are still not affordable in an experimental set-up, where the longest times that have been simulated are of order $\approx 10^3$ hopping units \cite{Aidelsburger_2020}, the results in Fig.~\ref{fig:XXZ_BlockDependence}~(c) provide us a hint of the kind of information learned by the neural network. We remind the reader that at such long timescales ($t_f \approx 10^8$), in an MBL phase, local degrees of freedom are frozen and the only relevant dynamics are induced by the dephasing mechanism producing a logarithmic slow propagation of entanglement \cite{Vznidarivc2008many,Bardarson2012unbounded,Serbyn2013universal}. As a result, we can fairly assume that our CNNs are able to distinguish an MBL phase from an Anderson insulator by learning dynamic correlations between the snapshots induced by the dephasing mechanism.

Finally, we would like to emphasize the important role of the time axis in the $3D$ structure of the input data. Figure~\ref{fig:XXZ_BlockDependence}~(d) shows the accuracy as a function of individual time $t$ of our CNNs, where the CNNs were trained using input data only from time $t$. Thus, the CNNs are not trained with the full dynamic range $t \in [0,t_f]$, but only on single time slices. To fairly compare the results, we want to give the same amount of information to our network and hence stack the $N_t=11$ slices of snapshots from the same time $t$ to form a $3D$ block. As expected, at short times $t\approx 1$, the neural network is not able to distinguish the two phases (accuracy $\approx 1/2$), since interactions do not play any relevant role yet.  Using larger times, a distinction is possible, though the performance is clearly worse in comparison to the performance of the CNNs trained with the full dynamic range. We conclude that the temporal correlation plays a major role in distinguishing the two phases.


\section{Network robustness} \label{sec:robustness}

Having tested the performance of our method, we now focus on the robustness of the CNNs. The goal of our work is to construct a neural network that can classify experimental data into MBL and Anderson insulator. Ideally, one would like to train a CNN using numerical data and then classify experimental data. This opens up the issue that the data from training and testing are originated from the same source, namely numerical simulations. Moreover, experimental data can have several forms of imperfections, i.e., the Hamiltonian's parameters are known only up to some precision. In the following two subsections, we show that the performance of our CNN is robust if tested on data produced with different Hamiltonian parameters. Then we train our CNN using Fock-space snapshots taken only from a small block of the system, allowing particle fluctuations. We show that information extracted from the small subsystem is already good enough to reliably distinguish the two phases~\footnote{This argument can work only if the localization length of of the system is larger than the subsystem.}.

\subsection{Robustness towards Hamiltonian perturbations}

In the following, we present an argument to provide evidence that our CNN approach is robust when tested on “imperfect” data sets and hence could be used in an experimental set-up. Figure~\ref{fig:XXZ_IntLength}~(a)  shows the fraction of correctly classified MBL \footnote{ fraction of correctly classified MBL $=  \frac{ \# \textit{MBL classified} }{\# \textit{ MBL}}$.} as a function of interaction strength $V\in [10^{-5}, 1]$ for systems of size $L=16$. Importantly, the CNNs have been trained using only the non-interacting samples ($V=0$) and the interacting ones with $V=1$. Moreover, Fig.~\ref{fig:XXZ_IntLength}~(a) also shows the results for different final times $t_f \in \{10^3, 10^6, 10^9\}$. A few considerations are in order. As expected, the time scale $t_f$ plays an important role. In agreement with the results in Fig.~\ref{fig:XXZ_BlockDependence}~(c), the accuracy increases with $t_f$, since the network is trained in a longer dynamic range. As expected, if $t_f \ll V^{-1}$ the performance is poor, since interactions have not shown their effects yet. This is manifested in the plateau of the blue curve ($t_f = 10^3$) that exists only until $t_f \cdot V \approx 10 $ (dashed vertical line in Fig.~\ref{fig:XXZ_IntLength}~(a)), where the accuracy starts to increase.  In Sec.~\ref{subsec:ComprehendHubbard}, we will see in detail that features like the logarithmic growth of the entanglement entropy $S(t) \sim \log{tV}$ are consistent with the information extracted by the networks. Thus, the rise of accuracy at $\sim t_f \cdot V^{-1}$ could be explained by the onset of the information growth~\cite{Serbyn2013universal}. Importantly, the performance of our CCN is robust if we moderately perturb the Hamiltonian's parameters, as can be seen for interaction strengths close to the value used for training ($V=1$). 

\subsection{Robustness towards subsystem classifications}
\label{sec:RobustnessSubsystems}
\begin{figure}[h!]
\includegraphics[width=1.\columnwidth]{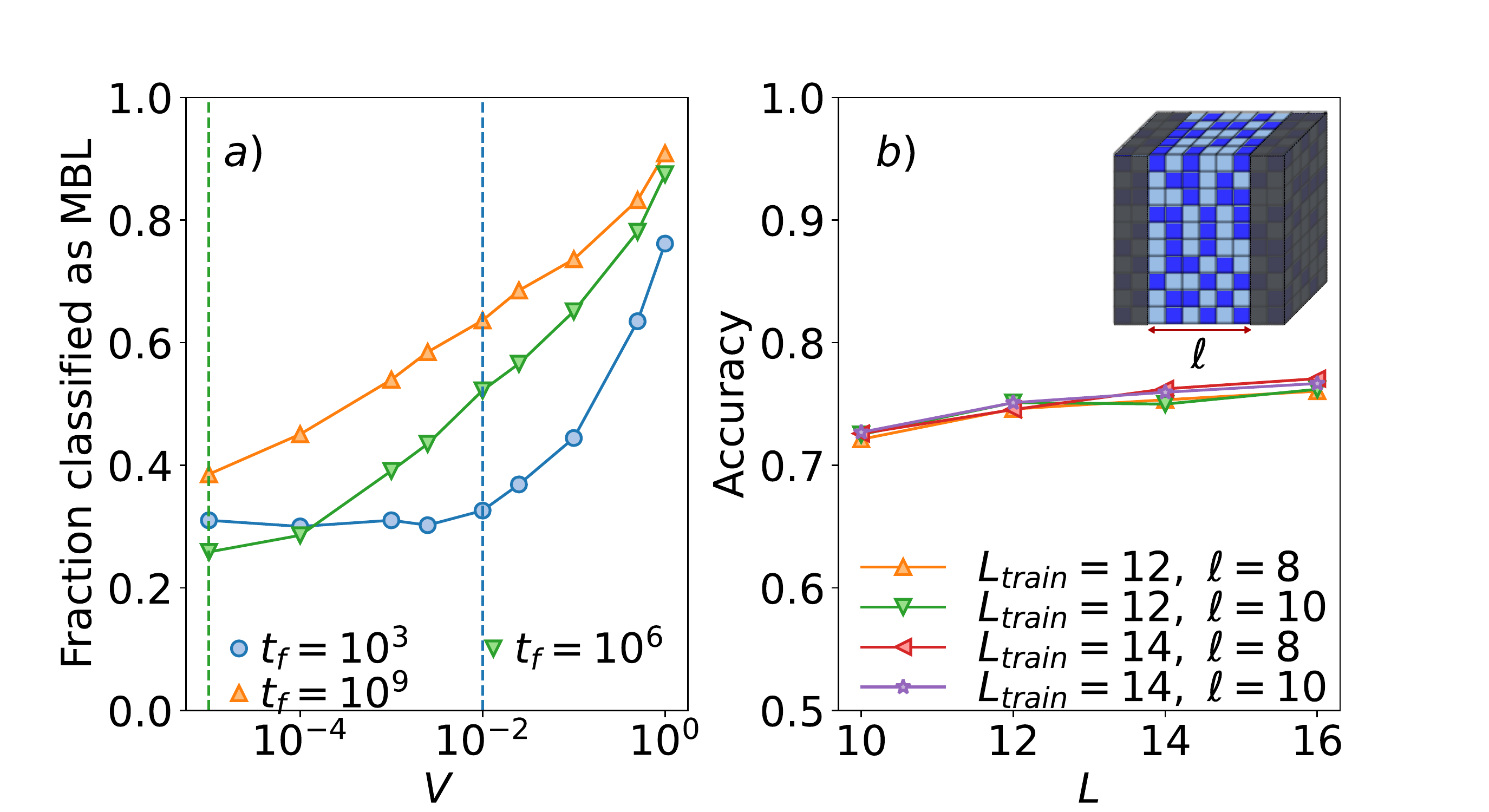}
\caption{(a) Performance of the CNNs on classifying data with $V \in [10^{-5},1]$, where the CNNs were only trained on different interaction strengths $V \in \{0,1\}$. Different curves represent the performance of the CNNs with different final times $t_f$. (b) Average accuracy for CNNs trained on sub-block snapshots of length $\ell$ in a system of system size $L_{\text{train}} \ge \ell$ and tested
on sub-block snapshots always of length $\ell$, but in a system of size $L$. }
\label{fig:XXZ_IntLength}
\end{figure}

Now we focus on the question of whether our method is stable if the CNN is trained using only Fock-state snapshots from a subsystem. The idea is to use cut-out techniques by training the CNN with Fock-state snapshots of length $\ell$ taken from a small subsystem ($n_{L_{\text{train}}/2-\ell/2},\cdots, n_{L_{\text{train}}/2+\ell/2}$) with $n_i \in\{0,1\}$ and $L_{\text{train}}$ the length of the system used to generate the snapshot blocks, see Fig.~\ref{fig:XXZ_IntLength}~(b)~\footnote{ Additionally, we exploit the fact that we can produce more cut-outs from larger systems, simply by using subsystems ($n_{0},\cdots, n_{\ell-1}$) for the first cut-out, then ($n_{1},\cdots, n_{\ell}$) for the second and so on until we reach the end of the chain ($n_{L-\ell},\cdots, n_{L-1}$). In testing, all cut-outs of one snapshot block are classified and labeled as one of the two phases. After going through all cut-outs of one snapshot block of length $L$, one assigns the category to the whole snapshot block which was ascribed to a majority of its cut-outs. We call this procedure a voting mechanism, since each classification of cut-outs gets one vote, and the majority vote decides which label is assigned to the whole system.}. 
Hence the CNN learns how to make a distinction while only seeing a small subsystem of length $\ell$ out of the entire system of length  $L_{\text{train}}$. 
\mbox{Figure~\ref{fig:XXZ_IntLength}~(b)} shows the test accuracy for several combinations of $\ell$, $L_{\text{train}}$ and the size of the system used to test the networks $L$. Remarkably, the performance of the network shows almost no dependence on $L_{\text{train}}$, $\ell$ and, in agreement with the results in Fig.~\ref{fig:XXZ_BlockDependence}~(b), the accuracy increases with $L$. This increase could be explained by the decreases of finite size effects with increasing $L$. The stability on $L_{\text{train}}$ and $\ell$ is a direct consequence of the fact that we are at strong disorder and therefore the localization length $\xi_{\text{loc}}$ is shorter than the dimension of the sub-system $\ell$. As a result, one might train the CNN using finite-size numerics and then use the network to classify data from experiments, which are usually done on larger system sizes. In particular, sub-block Fock-space snapshots have no particle number conservation, therefore we also tested the efficiency of our method in the case of a fluctuating number of particles.   

In summary, we have found that the CNN is stable to small perturbations of the Hamiltonian's parameters. Moreover, the CNN can have a high performance even if trained using only Fock-state snapshots taken from a smaller subsystem. As a consequence, only measurements in a finite portion of the system are needed to distinguish the two phases. Secondly, this opens the possibility to train our CNN using exact diagonalization techniques, which are limited to small system sizes.

\section{What does the neural network learn?} \label{subsec:ComprehendHubbard}

Understanding how a neural network makes a distinction and on which patterns it focuses is usually a challenging task \cite{Goodfellow2016DLBook}. The aim of this section is to evaluate which information the neural network uses to distinguish the two phases. 

Several methods have been proposed to solve this issue which are based on the examination of the kernels of the convolutional layer. For example, Schindler $et\ al.$~\cite{Schindler2017Probing} used the so-called dreaming mechanism~\cite{Mordvintsev2015Dreaming}, where a pre-trained network modifies random input data until this data would be classified in one of the two phases. By doing so, the network dreams about new data, meaning it produces new data that has the features of the phases which are critical for the classification. 

Here, we take the more pragmatic approach of taking a closer look at the correctly classified and misclassified samples. In Sec.~\ref{sec:robustness}, we have provided evidence that the neural network might learn features connected to the interaction-induced dephasing mechanism and consequentially to the information propagation through the system. Thus, motivated by the last observation we compare the dynamics of the correctly and wrongly classified samples using two different probes to distinguish the two phases. First, we compute the time evolution of the bipartite half-chain entanglement entropy 
\begin{equation}S(t) = - \text{Tr}[\rho_{L/2} \log{\rho_{L/2}}],
\end{equation} 
where $\rho_{L/2}(t)$ is the half-chain reduced density matrix of the evolved states $|\psi (t) \rangle $. In an MBL phase, after an initially short ballistic propagation,  $S(t)$ spreads logarithmically slow in time and finally reaches a non-thermal volume law steady-state for asymptotically long times ($S(\infty) \sim O(L)$). Instead, in an Anderson insulator phase entanglement does not propagate, and thus after the initial transient propagation, $S(t)$ saturates to an area law value ($S(\infty) \sim O(L^0)$). The second dynamic indicator we evaluate is the particle number fluctuation 
\begin{equation}
\Delta \mathcal{N}^2(t) = \langle O^2(t) \rangle -  \langle O(t) \rangle^2, \quad  O = \sum_{j=1}^{L/2} n_j. 
\end{equation}
In an interacting localized phase, $\Delta \mathcal{N}^2(t)$ has an extremely slow propagation over several orders of magnitude in time, which is consistent with $\Delta \mathcal{N}^2(t) \propto \log{\log{t}}$, unlike in an Anderson localized phase. Although it is debated if the found propagation is only transient or persistent for asymptotically long times~\cite{Bar_Lev_2020, Jesko_2021}, $\Delta \mathcal{N}^2(t)$ still remains a useful and experimentally accessible probe to distinguish the two phases~\cite{Smith2016QSimulator} at finite time-scales. 

During the time evolution, we compute both the entanglement entropy $S(t)$ and the particle fluctuation $\Delta \mathcal{N}^2(t)$ and simultaneously sample the Fock-state snapshots $\underline{n} = (n_1,..., n_L)$ from the probability distribution $P_{\underline{n}} (t)$ in Eq.~\ref{eq:Probability}. Finally, we test our CNN on this data, and in order to understand which features the CNN learns,  we take a closer look at $S(t)$ and $\Delta \mathcal{N}^2(t)$ for the samples that have been correctly and wrongly classified. 

\begin{figure}[h!]
\includegraphics[width=1.\columnwidth,trim={.5cm 4cm 1cm 1cm},clip]{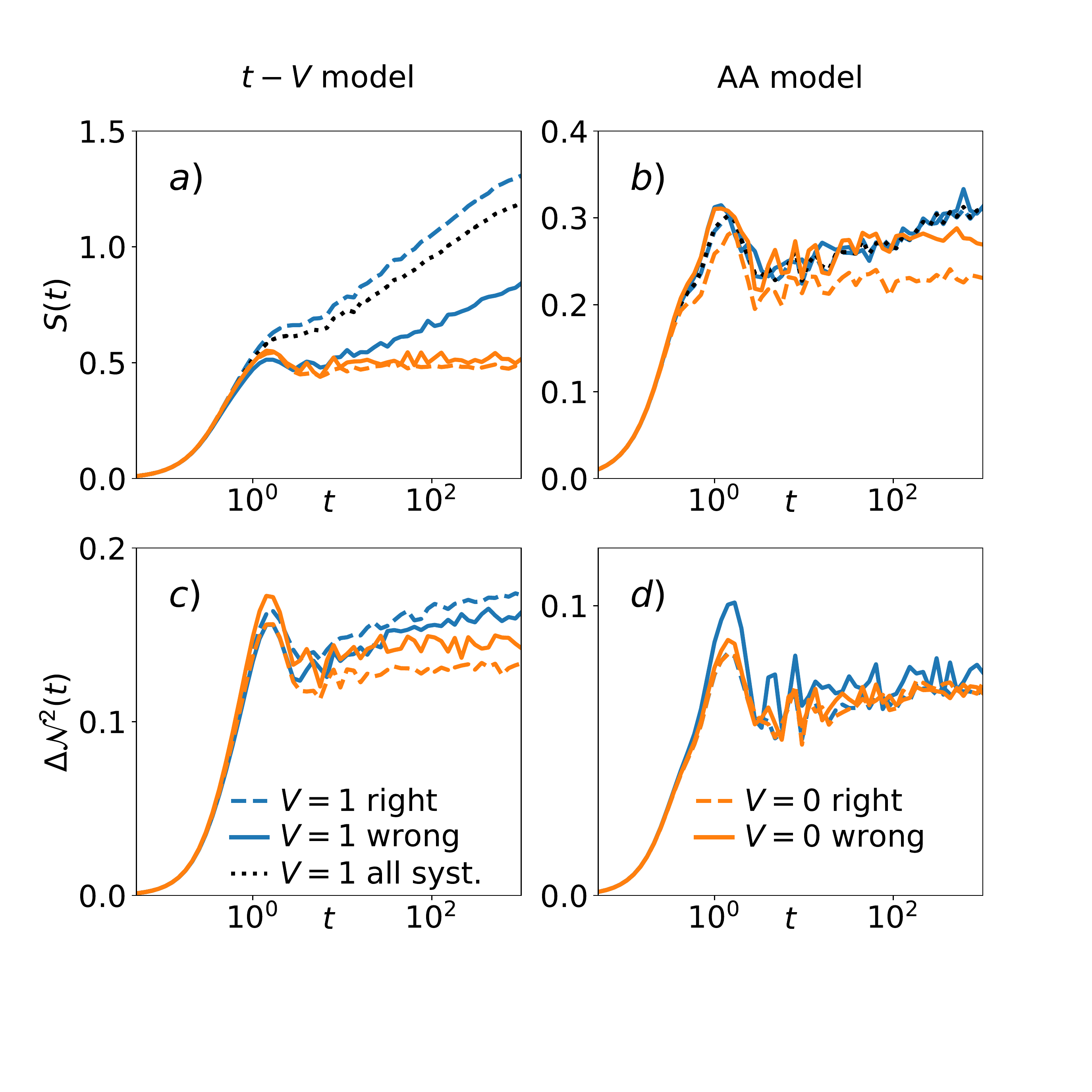}
\caption{ (a),(c) show the growth of the entanglement entropy $S(t)$ and particle fluctuations $\Delta \mathcal{N}^2(t)$ averaged separately over the correctly and wrongly classified Anderson ($V=0$) and MBL ($V=1$) samples for the disordered $t-V$ model. (b),(d) show the same information as in the adjacent panels, but for the AA model with quasi-periodic potential.}
\label{fig:XXZAubry_Understanding}
\end{figure}

Figures~\ref{fig:XXZAubry_Understanding}~(a), (c) show the entanglement entropy $S(t)$ and the particle number fluctuation $\Delta \mathcal{N}^2(t)$ averaged separately over correctly and wrongly classified samples (dashed lines), respectively. For the interacting case, $t-V$ model with $V=1$, a clear separation between the correctly and wrongly classified snapshots is visible. Crucially, the samples which the network wrongly classified as Anderson localized exhibit a late onset and a slower propagation of information, thus affirming our main assumption. 

The situation changes drastically when we consider the interacting model having quasi-periodic potential (AA model), defined in Sec.~\ref{sec:Model}~\cite{Shankar_2012,Lim_2017,Sarang_2021}. In order to have a fair comparison to the $t-V$ model, we train our CNNs with the same $3D$ structure of the input data $\# \text{Snapshots} \times L \times N_t$ with $\# \text{Snapshots} = 30$, $L=16$, $N_t=11$, final time $t_f = 10^3$, and $W \in \{6,7,8\}$.
 \begin{figure}[h]
\includegraphics[width=0.7\columnwidth]{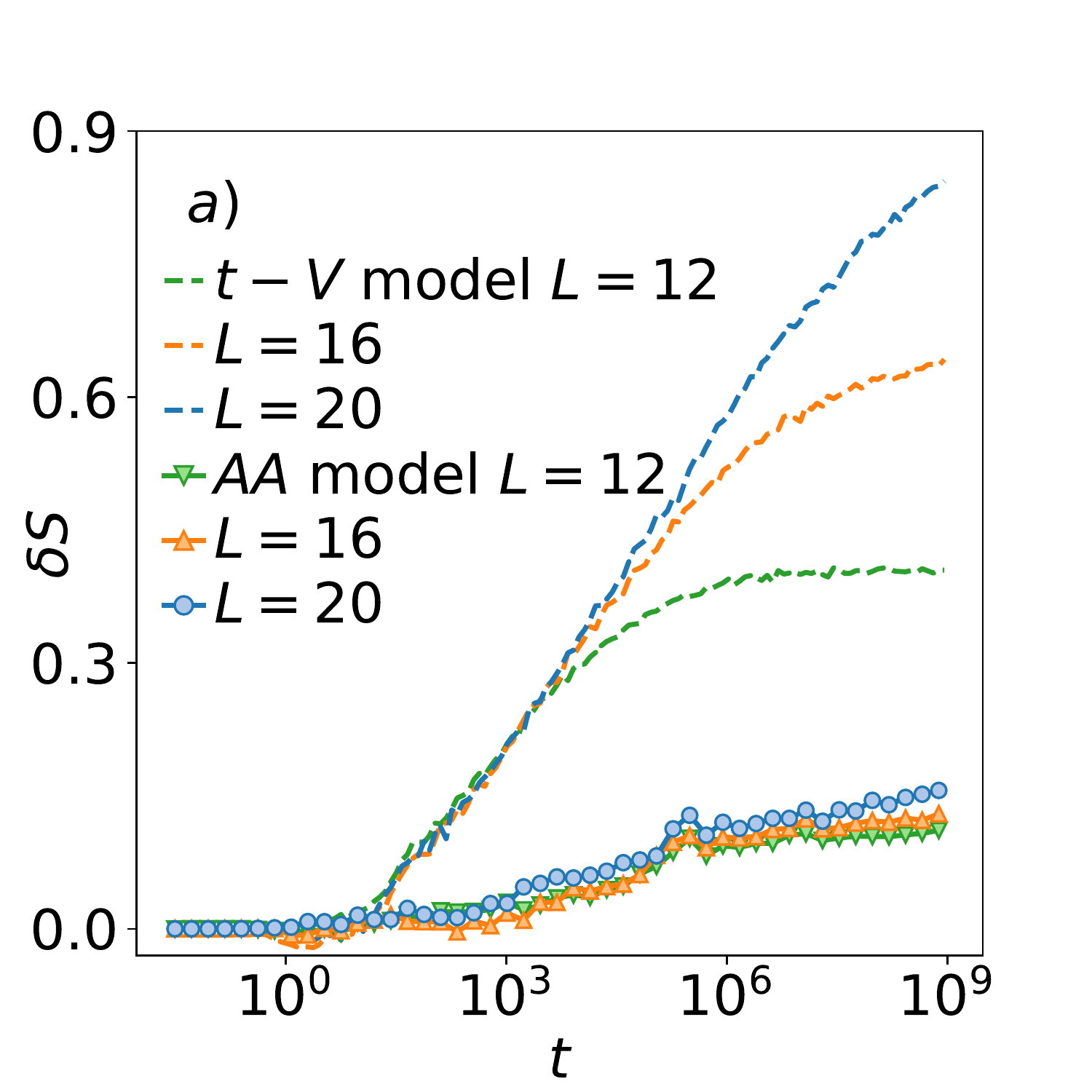}
\caption{ (a) shows the entanglement entropy after subtracting the non-interacting value $V=0$. The dynamics is computed using the $\ell$-bit Hamiltonian $H^{\text{eff}}$ with $W=6$ and $V=0.1$, for the $t-V$ (dashed line) and the AA (solid line) models.}
\label{fig:DeltaS}
\end{figure}

In the case of the $t-V$ model, the CNN reaches an accuracy of $\approx 76\%$, while for the AA model the network is on average only able to classify $\approx 66\%$ of the samples correctly (see Appendix). The difficulty for the CNN to distinguish the two phases for the AA model is connected to an atypically slow information propagation through the system. Figure~\ref{fig:XXZAubry_Understanding}~(b) and (d) show $S(t)$ and $\Delta \mathcal{N}^2(t)$ averaged only over correctly and wrongly classified samples, respectively.  As one can observe, in the AA model the information propagation is much slower than in the case with the random potential (see Figs.~\ref{fig:XXZAubry_Understanding}~(a), (c)).  Particularly, $S(t)$ has only a growth of $\approx 30 \%$ at timescale $t_f = 10^3$, compared to the non-interacting case, and $\Delta \mathcal{N}^2(t)$ does not show any significant growth. As a consequence, our CNNs have difficulties in distinguishing the two phases.

This anomalous propagation is traceable to the fractal nature of the single-particle spectrum of the AA model~\cite{casati_chirikov_1995}. Indeed, it is known that that the energy spectrum of the AA model is multifractal and thus its spectrum hosts quasi-degenerate single-particle energy~\cite{casati_chirikov_1995}. As a consequence, these quasi-degenerate frozen degrees of freedom with energies $\epsilon_x$ and $\epsilon_y$ will need a time proportional to $(\epsilon_x - \epsilon_y)^{-1}$ to dephase. We test this idea by studying the EE growth for a minimal-model constructed from $H$ in Eq~\ref{eq:Hamiltonian} using perturbation theory in the interaction strength $V$. In the limit of weak interactions $V\ll 1$ and strong disorder $W\gg 1$, $H$ can be approximated as $H^{\text{eff}} = \sum_l \epsilon_l \eta_l^\dagger \eta_l + V\sum_{l,m} \mathcal{B}_{lm} \eta_l^\dagger \eta_l\eta_m^\dagger \eta_m$, where $\eta_l^\dagger (\eta_l)$ is the creation (annihilation) operator for the single-particle orbital $\phi_l(x)$ at energy $\epsilon_l$, and $\mathcal{B}_{lm}$ can be found by using perturbation theory (for details see Ref.~\onlinecite{DeTomasi2019Efficient}). In this limit, 
$H^{\text{eff}}$ represents the paradigmatic model known as $\ell$-bits Hamiltonian~\cite{Huse_2014_fully, Ros2015Integrals, Imbrie_2017}, describing an MBL phase at strong disorder. 

Figure~\ref{fig:DeltaS} shows $\delta S$, which is the entanglement entropy after subtracting the non-interacting values ($V=0$). $\delta S$ is computed using $H^{\text{eff}}$ constructed from the $t-V$ model (dashed lines) and the AA model (solid lines). As one can observe and in agreement with the results in Fig.~\ref{fig:XXZAubry_Understanding}, in the MBL phase for the AA model $S(t)$ has an anomalous slow propagation if compared to the $t-V$ model case. We checked numerically that the dephasing couplings $\mathcal{B}_{l,m}$ in $H^{\text{eff}}$ do not present significant differences between the two models. Thus, the important difference between the two models is only given by the single-particle energies $\{\epsilon_l\}$. Indeed, in the case of the $t-V$ model the $\{\epsilon_l\}$ are Poissonian distributed and their density of states can be approximated with a box function at strong disorder, whereas in the AA model the $\{\epsilon_l\}$ are almost-degenerate and the density of states is a Cantor set. As a result, this provides numerical evidence that the slower propagation of information for the AA model is due to the fractal nature of its single-particle spectrum~\cite{casati_chirikov_1995}.


\section{Conclusion} \label{sec:conclusion}

In this work, we addressed the question of how to distinguish an Anderson insulator from an MBL phase using snapshot data by formulating a method based on machine learning tools, i.e., CNNs. We trained a CNN using a $3D$ structure for the input data, which contains a fixed amount of space-time Fock-state snapshots. A particular focus was given to this $3D$ structure and on the fundamental importance of having dynamic/temporal information. Unlike the case of distinguishing an ergodic phase from a localized one, where only a fixed amount of snapshots is important~\cite{Bohrdt_2020}, here it is crucial to analyze snapshots at different times. Thus, this work provides a novel method to analyze accessible experimental data (e.g. from cold-atoms setups) and to separate the two localized phases. 

We benchmarked our method on the paradigmatic model exhibiting an MBL phase ($t-V$ model with quenched disorder) and showed that the CNN reaches accuracies of $\approx 80\%$ in distinguishing the two phases. We studied the stability of our method and provided evidence that it can be used to analyze real experimental data. Importantly, we found an upper bound to the number of snapshots needed in an experimental setup to distinguish the two phases. 

In order to understand what kind of features of the input data is used by the CNN to classify the two phases, we took a closer look at the dynamics of the entanglement entropy and the particle fluctuation. Both quantities are known to be dynamical indicators of an MBL phase. From this analysis, we provided evidence that the CNN makes use of the difference in the information propagation between an MBL phase and an Anderson localized one. In fact, we showed that the interacting samples misclassified as Anderson localized are characterized by an unusually slow entanglement propagation. 

Finally, we applied our method to a model with quasi-periodic potential (AA model). Like the $t-V$ model, this model has been used several times in cold-atoms experiments to study the MBL transition. In this case, the neural networks had difficulties when separating the phases, and the reached accuracy is significantly lower than in the case of fully random potentials. This is due to a quantitative slower propagation of information in the AA model compared to the $t-V$ model. Indeed, we showed that the MBL phase of the AA model has a much slower growth of entanglement and the particle fluctuation fails to be a dynamical indicator to distinguish the two phases. These results indicate that separating an MBL phase from an Anderson insulator in the case of quasi-periodic is potentially more challenging than for a disordered system. 

\begin{acknowledgments}
We would like to thank F. Meissen for his insightful comments.
We also express our gratitude to S. Bera, R. Hardman Carter, S. Kim, and S. Lechner for the critical reading of the manuscript. FP acknowledges support from the European Research Council (ERC) under the European Union’s Horizon 2020 research and innovation programme (grant agreement No. 771537). FP acknowledges the support of the Deutsche Forschungs-gemeinschaft (DFG, German Research Foundation) under Germany’s Excellence Strategy EXC-2111-390814868 and TRR 80.
\end{acknowledgments}
\appendix 
\section*{Appendix} \label{sec:appe}

\subsection{Network architecture and hyperparameters}

\begin{figure}
\centering
\vspace{1cm}
\def\svgwidth{1.\columnwidth}
\fontsize{6}{10}\selectfont
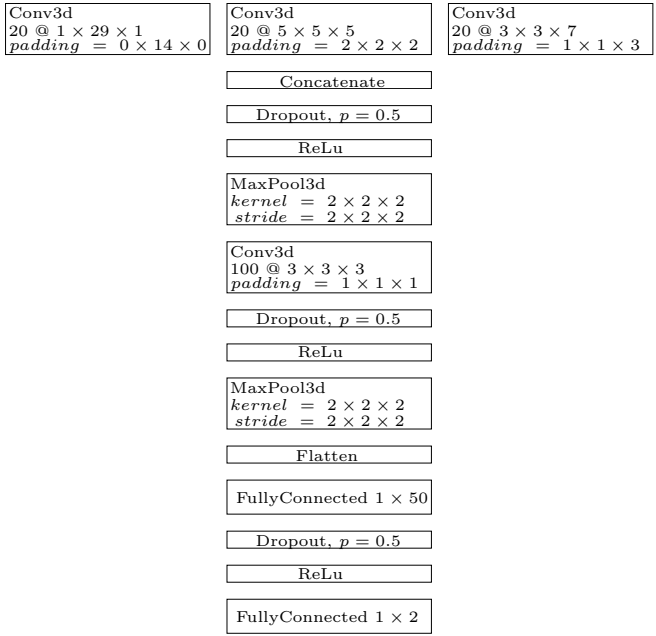
\caption{Schematic representation of the CNN. The first layer is an adaptation of an inception layer \cite{Szegedy2014Inception}, followed by a simple convolutional layer. The final classification is done by two fully connected layers. Two three-dimensional MaxPool layers shrink down the data size after the inception layer and after the convolutional layer. Nonlinearities are introduced by the rectified linear unit (ReLU). Dropout is included to increase the classification accuracy.}
\label{fig:NNsAppendix}
\end{figure}

The CNN used in this work has a rather simple architecture, which allows extracting complicated dynamic patterns from the $3D$ snapshot input. It consists of two convolution layers with $3D$ kernels and two fully connected layers.
Figure~\ref{fig:NNsAppendix} shows a schematic representation of the neural network. The first layer is an adaptation of the inception layer ~\cite{Szegedy2014Inception}, where three different kernels extract information from the input data. One kernel of size $(1 \times 29 \times 1)$ is able to see many snapshots of one site at a time, allowing it to compute averages over the snapshots. The second kernel of size $(5 \times 5 \times 5)$ gets information from different sites, snapshots and times. The third kernel of size $(3 \times 3 \times 7)$ is specially designed to extract dynamic information, hence it has access to $7$ points in time. All kernels have a stride equals to one and a padding is chosen to match the dimension of the output after the convolution layer to the dimension of the input (known as \textit{same padding}). A pooling layer reduces the amount of information, before the feature map is given to a second convolution layer with a kernel size of $(2 \times 2 \times 2)$, followed by another pooling layer. The last two layers are fully connected layers with $50$ and $2$ neurons to make the classification.\\
The optimizer used in training is Adam~\cite{Kingma2014Adam}. Note that we use a smaller learning rate of $10^{-4}$ since the input data heavily depends on the random collapse of the wavefunction 
The loss function we use is the cross-entropy loss, additional hyperparameters are listed in Tab.~\ref{tab:hyperparameters}. The CNNs were implemented in Pytorch~\cite{Paszke2019Pytorch}.

\begin{table}[h]
\centering
\begin{tabular}{l | l }
hyperparameter   & value \\
\hline
learning rate & $10^{-4}$  \\
learning rate decay &  $0.995$ \\
weight decay &  $0$ \\
number of epochs &  $150$ \\
\end{tabular}
\caption{Hyperparameters we use for implementation and training. Parameters not listed are the standard parameters proposed by Pytorch.}
\label{tab:hyperparameters}
\end{table}

The training was performed on GPUs supporting CUDA platform, namely a GeForce GTX 960, GeForce GTX 1050 Ti, and GeForce GTX 1650 with a minimum memory of $4$ GB.\\

\subsection{Quantum Fisher Information}

As in Sec.~\ref{subsec:ComprehendHubbard}, we use the quantum Fisher information (QFI) to compare the dynamics of the right and wrong classified samples. The QFI is defined by

\begin{equation}
    \mathcal{F}(t)=4 \left[ \langle \hat{\mathcal{O}}^2 \rangle -\langle \hat{\mathcal{O}} \rangle^2 \right], \;\; \hat{\mathcal{O}} = \sum_i (-1)^i \hat{n}_i.
\end{equation}

\begin{figure}[h!]
\includegraphics[width=1.\columnwidth,trim={1cm 2cm 1cm -2cm},clip]{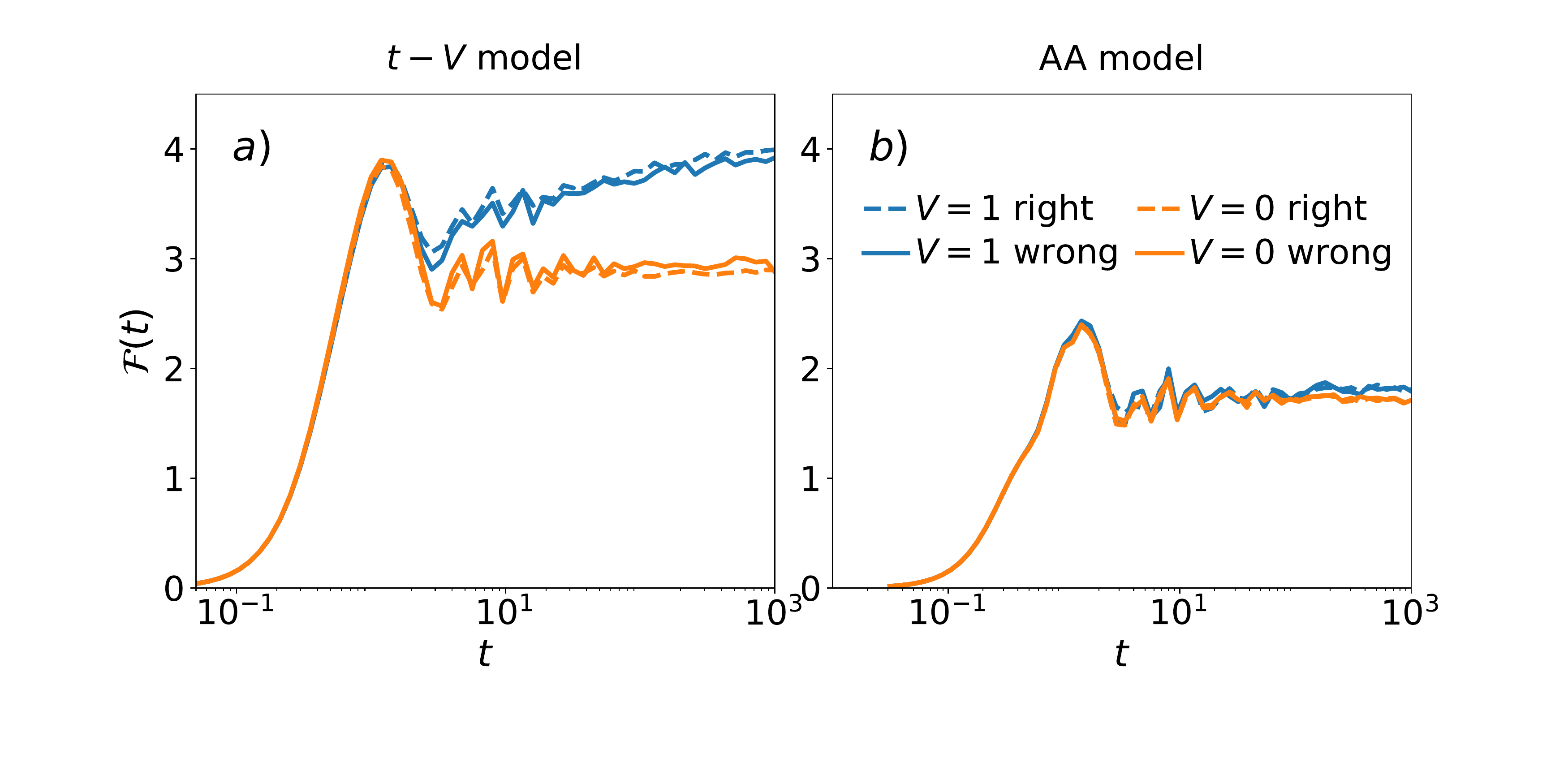}
\caption{Both panels show the QFI $\mathcal{F}(t)$ averaged separately over the correctly and wrongly classified MBL and Anderson insulating samples for the $t-V$ model (a) and the AA model (b).}
\label{fig:QFI_Appendix}
\end{figure}

In Fig.~\ref{fig:QFI_Appendix}~(a) a clear distinction between right and wrong classifications of the two phases with $V=0,1$ cannot be made when examining  $\mathcal{F}(t)$, indicating that the network does not focus on patterns that are comparable to $\mathcal{F}(t)$. For the quasi-periodic AA model in (b), we see that $\mathcal{F}(t)$ is not a good quantity even to distinguish $V=0$ and $V=1$ since it exhibits no different dynamics for small times $t<10^3$.

\subsection{Network performance and robustness for the AA model}
In this section, we test the robustness of the CNNs and their general performance to the case with the quasi-periodic potential (AA model), just as we did in chapter~\ref{sec:NetworkPerformance} and~\ref{sec:robustness} for the $t-V$ model. We find out that the neural network performs worse in the case of quasi-periodic potentials.
 \begin{figure}[b]
\includegraphics[width=1\columnwidth,trim={0cm 0cm 0cm 0cm},clip]{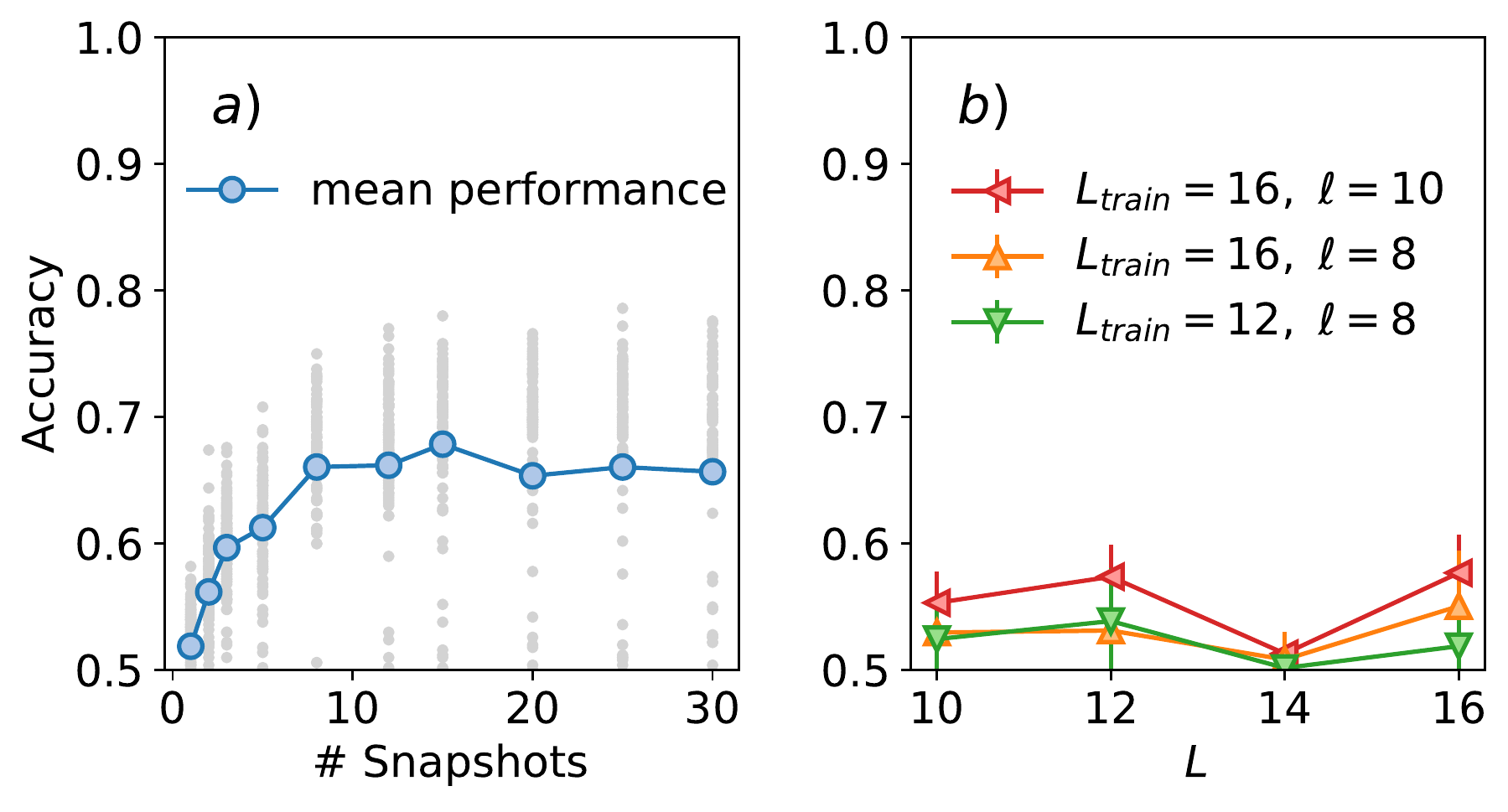}
\caption{(a) Dependence of the network performance on $\# \text{Snapshots}$ in the AA model. The blue line indicates the averaged performance over 50 networks, grey dots mark the performance of single CNNs. (b)  Average accuracy for CNNs trained on sub-block snapshots of length $\ell$ in a system of size $L_{\text{train}} \ge \ell$ and tested on sub-block snapshots always of length $\ell$, but in a system of size $L$ in the AA model.}
\label{fig:AA_Appendix}
\end{figure}

In Fig.~\ref{fig:AA_Appendix}~(a), we test the dependence of the network performance on the shape of the snapshot blocks in the AA model, comparable to Fig.~\ref{fig:XXZ_BlockDependence}. The classification accuracy rises with rising $\# \text{Snapshots}$ up to a saturation value of about $65\%$, which is roughly $10\%$ less than in the $t-V$ model. The snapshot blocks are taken from systems with $L=16$, $t_f=10^3$, the result is averaged over $50$ networks, grey dots show the performance of single CNNs.

In Fig.~\ref{fig:AA_Appendix}~(b) we also test the robustness of our network when we change the chain length $L$ in the AA model. Therefore, we use cut-offs of length $\ell$ defined in Sec.~\ref{sec:robustness}. For the $t-V$ model we saw good classification results when we trained our model using smaller blocks of length $\ell$, see Fig.~\ref{fig:XXZ_IntLength}~(b). However, the very same idea and network architecture fail for the AA model, producing classification accuracies $<60\%$ which is only slightly better than tossing a fair coin.

Hence, we conclude that the dynamic patterns the CNN can extract from the $3D$ snapshot blocks are less apparent in the quasi-periodic AA model than in the $t-V$ model.

\bibliography{NNs_MBL_bib}

\begin{thebibliography}{81}%
\makeatletter
\providecommand \@ifxundefined [1]{%
 \@ifx{#1\undefined}
}%
\providecommand \@ifnum [1]{%
 \ifnum #1\expandafter \@firstoftwo
 \else \expandafter \@secondoftwo
 \fi
}%
\providecommand \@ifx [1]{%
 \ifx #1\expandafter \@firstoftwo
 \else \expandafter \@secondoftwo
 \fi
}%
\providecommand \natexlab [1]{#1}%
\providecommand \enquote  [1]{``#1''}%
\providecommand \bibnamefont  [1]{#1}%
\providecommand \bibfnamefont [1]{#1}%
\providecommand \citenamefont [1]{#1}%
\providecommand \href@noop [0]{\@secondoftwo}%
\providecommand \href [0]{\begingroup \@sanitize@url \@href}%
\providecommand \@href[1]{\@@startlink{#1}\@@href}%
\providecommand \@@href[1]{\endgroup#1\@@endlink}%
\providecommand \@sanitize@url [0]{\catcode `\\12\catcode `\$12\catcode
  `\&12\catcode `\#12\catcode `\^12\catcode `\_12\catcode `\%12\relax}%
\providecommand \@@startlink[1]{}%
\providecommand \@@endlink[0]{}%
\providecommand \url  [0]{\begingroup\@sanitize@url \@url }%
\providecommand \@url [1]{\endgroup\@href {#1}{\urlprefix }}%
\providecommand \urlprefix  [0]{URL }%
\providecommand \Eprint [0]{\href }%
\providecommand \doibase [0]{https://doi.org/}%
\providecommand \selectlanguage [0]{\@gobble}%
\providecommand \bibinfo  [0]{\@secondoftwo}%
\providecommand \bibfield  [0]{\@secondoftwo}%
\providecommand \translation [1]{[#1]}%
\providecommand \BibitemOpen [0]{}%
\providecommand \bibitemStop [0]{}%
\providecommand \bibitemNoStop [0]{.\EOS\space}%
\providecommand \EOS [0]{\spacefactor3000\relax}%
\providecommand \BibitemShut  [1]{\csname bibitem#1\endcsname}%
\let\auto@bib@innerbib\@empty
\bibitem [{\citenamefont {Gross}\ and\ \citenamefont {Bloch}(2017)}]{Gross995}%
  \BibitemOpen
  \bibfield  {author} {\bibinfo {author} {\bibfnamefont {C.}~\bibnamefont
  {Gross}}\ and\ \bibinfo {author} {\bibfnamefont {I.}~\bibnamefont {Bloch}},\
  }\bibfield  {title} {\bibinfo {title} {Quantum simulations with ultracold
  atoms in optical lattices},\ }\href {https://doi.org/10.1126/science.aal3837}
  {\bibfield  {journal} {\bibinfo  {journal} {Science}\ }\textbf {\bibinfo
  {volume} {357}},\ \bibinfo {pages} {995} (\bibinfo {year} {2017})},\ \Eprint
  {https://arxiv.org/abs/https://science.sciencemag.org/content/357/6355/995.full.pdf}
  {https://science.sciencemag.org/content/357/6355/995.full.pdf} \BibitemShut
  {NoStop}%
\bibitem [{\citenamefont {Monroe}\ \emph {et~al.}(2021)\citenamefont {Monroe},
  \citenamefont {Campbell}, \citenamefont {Duan}, \citenamefont {Gong},
  \citenamefont {Gorshkov}, \citenamefont {Hess}, \citenamefont {Islam},
  \citenamefont {Kim}, \citenamefont {Linke}, \citenamefont {Pagano},
  \citenamefont {Richerme}, \citenamefont {Senko},\ and\ \citenamefont
  {Yao}}]{Monroe_2021}%
  \BibitemOpen
  \bibfield  {author} {\bibinfo {author} {\bibfnamefont {C.}~\bibnamefont
  {Monroe}}, \bibinfo {author} {\bibfnamefont {W.~C.}\ \bibnamefont
  {Campbell}}, \bibinfo {author} {\bibfnamefont {L.-M.}\ \bibnamefont {Duan}},
  \bibinfo {author} {\bibfnamefont {Z.-X.}\ \bibnamefont {Gong}}, \bibinfo
  {author} {\bibfnamefont {A.~V.}\ \bibnamefont {Gorshkov}}, \bibinfo {author}
  {\bibfnamefont {P.~W.}\ \bibnamefont {Hess}}, \bibinfo {author}
  {\bibfnamefont {R.}~\bibnamefont {Islam}}, \bibinfo {author} {\bibfnamefont
  {K.}~\bibnamefont {Kim}}, \bibinfo {author} {\bibfnamefont {N.~M.}\
  \bibnamefont {Linke}}, \bibinfo {author} {\bibfnamefont {G.}~\bibnamefont
  {Pagano}}, \bibinfo {author} {\bibfnamefont {P.}~\bibnamefont {Richerme}},
  \bibinfo {author} {\bibfnamefont {C.}~\bibnamefont {Senko}},\ and\ \bibinfo
  {author} {\bibfnamefont {N.~Y.}\ \bibnamefont {Yao}},\ }\bibfield  {title}
  {\bibinfo {title} {Programmable quantum simulations of spin systems with
  trapped ions},\ }\href {https://doi.org/10.1103/RevModPhys.93.025001}
  {\bibfield  {journal} {\bibinfo  {journal} {Rev. Mod. Phys.}\ }\textbf
  {\bibinfo {volume} {93}},\ \bibinfo {pages} {025001} (\bibinfo {year}
  {2021})}\BibitemShut {NoStop}%
\bibitem [{\citenamefont {Blatt}\ and\ \citenamefont {Roos}(2012)}]{Blatt2012}%
  \BibitemOpen
  \bibfield  {author} {\bibinfo {author} {\bibfnamefont {R.}~\bibnamefont
  {Blatt}}\ and\ \bibinfo {author} {\bibfnamefont {C.~F.}\ \bibnamefont
  {Roos}},\ }\bibfield  {title} {\bibinfo {title} {Quantum simulations with
  trapped ions},\ }\href {https://doi.org/10.1038/nphys2252} {\bibfield
  {journal} {\bibinfo  {journal} {Nature Physics}\ }\textbf {\bibinfo {volume}
  {8}},\ \bibinfo {pages} {277} (\bibinfo {year} {2012})}\BibitemShut {NoStop}%
\bibitem [{\citenamefont {Kjaergaard}\ \emph {et~al.}(2020)\citenamefont
  {Kjaergaard}, \citenamefont {Schwartz}, \citenamefont {Braumüller},
  \citenamefont {Krantz}, \citenamefont {Wang}, \citenamefont {Gustavsson},\
  and\ \citenamefont {Oliver}}]{Kjaergaard_2020}%
  \BibitemOpen
  \bibfield  {author} {\bibinfo {author} {\bibfnamefont {M.}~\bibnamefont
  {Kjaergaard}}, \bibinfo {author} {\bibfnamefont {M.~E.}\ \bibnamefont
  {Schwartz}}, \bibinfo {author} {\bibfnamefont {J.}~\bibnamefont
  {Braumüller}}, \bibinfo {author} {\bibfnamefont {P.}~\bibnamefont {Krantz}},
  \bibinfo {author} {\bibfnamefont {J.~I.-J.}\ \bibnamefont {Wang}}, \bibinfo
  {author} {\bibfnamefont {S.}~\bibnamefont {Gustavsson}},\ and\ \bibinfo
  {author} {\bibfnamefont {W.~D.}\ \bibnamefont {Oliver}},\ }\bibfield  {title}
  {\bibinfo {title} {Superconducting qubits: Current state of play},\ }\href
  {https://doi.org/10.1146/annurev-conmatphys-031119-050605} {\bibfield
  {journal} {\bibinfo  {journal} {Annual Review of Condensed Matter Physics}\
  }\textbf {\bibinfo {volume} {11}},\ \bibinfo {pages} {369} (\bibinfo {year}
  {2020})},\ \Eprint
  {https://arxiv.org/abs/https://doi.org/10.1146/annurev-conmatphys-031119-050605}
  {https://doi.org/10.1146/annurev-conmatphys-031119-050605} \BibitemShut
  {NoStop}%
\bibitem [{\citenamefont {Bloch}\ \emph {et~al.}(2008)\citenamefont {Bloch},
  \citenamefont {Dalibard},\ and\ \citenamefont {Zwerger}}]{Bloch_review_2008}%
  \BibitemOpen
  \bibfield  {author} {\bibinfo {author} {\bibfnamefont {I.}~\bibnamefont
  {Bloch}}, \bibinfo {author} {\bibfnamefont {J.}~\bibnamefont {Dalibard}},\
  and\ \bibinfo {author} {\bibfnamefont {W.}~\bibnamefont {Zwerger}},\
  }\bibfield  {title} {\bibinfo {title} {Many-body physics with ultracold
  gases},\ }\href {https://doi.org/10.1103/RevModPhys.80.885} {\bibfield
  {journal} {\bibinfo  {journal} {Rev. Mod. Phys.}\ }\textbf {\bibinfo {volume}
  {80}},\ \bibinfo {pages} {885} (\bibinfo {year} {2008})}\BibitemShut
  {NoStop}%
\bibitem [{\citenamefont {Schreiber}\ \emph {et~al.}(2015)\citenamefont
  {Schreiber}, \citenamefont {Hodgman}, \citenamefont {Bordia}, \citenamefont
  {L{\"u}schen}, \citenamefont {Fischer}, \citenamefont {Vosk}, \citenamefont
  {Altman}, \citenamefont {Schneider},\ and\ \citenamefont
  {Bloch}}]{Schreiber2015Coldatoms}%
  \BibitemOpen
  \bibfield  {author} {\bibinfo {author} {\bibfnamefont {M.}~\bibnamefont
  {Schreiber}}, \bibinfo {author} {\bibfnamefont {S.~S.}\ \bibnamefont
  {Hodgman}}, \bibinfo {author} {\bibfnamefont {P.}~\bibnamefont {Bordia}},
  \bibinfo {author} {\bibfnamefont {H.~P.}\ \bibnamefont {L{\"u}schen}},
  \bibinfo {author} {\bibfnamefont {M.~H.}\ \bibnamefont {Fischer}}, \bibinfo
  {author} {\bibfnamefont {R.}~\bibnamefont {Vosk}}, \bibinfo {author}
  {\bibfnamefont {E.}~\bibnamefont {Altman}}, \bibinfo {author} {\bibfnamefont
  {U.}~\bibnamefont {Schneider}},\ and\ \bibinfo {author} {\bibfnamefont
  {I.}~\bibnamefont {Bloch}},\ }\bibfield  {title} {\bibinfo {title}
  {Observation of many-body localization of interacting fermions in a
  quasirandom optical lattice},\ }\href
  {https://doi.org/10.1126/science.aaa7432} {\bibfield  {journal} {\bibinfo
  {journal} {Science}\ }\textbf {\bibinfo {volume} {349}},\ \bibinfo {pages}
  {842} (\bibinfo {year} {2015})}\BibitemShut {NoStop}%
\bibitem [{\citenamefont {{Smith}}\ \emph {et~al.}(2016)\citenamefont
  {{Smith}}, \citenamefont {{Lee}}, \citenamefont {{Richerme}}, \citenamefont
  {{Neyenhuis}}, \citenamefont {{Hess}}, \citenamefont {{Hauke}}, \citenamefont
  {{Heyl}}, \citenamefont {{Huse}},\ and\ \citenamefont
  {{Monroe}}}]{Smith2016QSimulator}%
  \BibitemOpen
  \bibfield  {author} {\bibinfo {author} {\bibfnamefont {J.}~\bibnamefont
  {{Smith}}}, \bibinfo {author} {\bibfnamefont {A.}~\bibnamefont {{Lee}}},
  \bibinfo {author} {\bibfnamefont {P.}~\bibnamefont {{Richerme}}}, \bibinfo
  {author} {\bibfnamefont {B.}~\bibnamefont {{Neyenhuis}}}, \bibinfo {author}
  {\bibfnamefont {P.~W.}\ \bibnamefont {{Hess}}}, \bibinfo {author}
  {\bibfnamefont {P.}~\bibnamefont {{Hauke}}}, \bibinfo {author} {\bibfnamefont
  {M.}~\bibnamefont {{Heyl}}}, \bibinfo {author} {\bibfnamefont {D.~A.}\
  \bibnamefont {{Huse}}},\ and\ \bibinfo {author} {\bibfnamefont
  {C.}~\bibnamefont {{Monroe}}},\ }\bibfield  {title} {\bibinfo {title}
  {{Many-body localization in a quantum simulator with programmable random
  disorder}},\ }\href
  {http://www.nature.com/nphys/journal/v12/n10/full/nphys3783.html} {\bibfield
  {journal} {\bibinfo  {journal} {Nat. Phys.}\ }\textbf {\bibinfo {volume}
  {12}},\ \bibinfo {pages} {907} (\bibinfo {year} {2016})}\BibitemShut
  {NoStop}%
\bibitem [{\citenamefont {Choi}\ \emph {et~al.}(2016)\citenamefont {Choi},
  \citenamefont {Hild}, \citenamefont {Zeiher}, \citenamefont {Schau{\ss}},
  \citenamefont {Rubio-Abadal}, \citenamefont {Yefsah}, \citenamefont
  {Khemani}, \citenamefont {Huse}, \citenamefont {Bloch},\ and\ \citenamefont
  {Gross}}]{Choi2016Coldatoms}%
  \BibitemOpen
  \bibfield  {author} {\bibinfo {author} {\bibfnamefont {J.-y.}\ \bibnamefont
  {Choi}}, \bibinfo {author} {\bibfnamefont {S.}~\bibnamefont {Hild}}, \bibinfo
  {author} {\bibfnamefont {J.}~\bibnamefont {Zeiher}}, \bibinfo {author}
  {\bibfnamefont {P.}~\bibnamefont {Schau{\ss}}}, \bibinfo {author}
  {\bibfnamefont {A.}~\bibnamefont {Rubio-Abadal}}, \bibinfo {author}
  {\bibfnamefont {T.}~\bibnamefont {Yefsah}}, \bibinfo {author} {\bibfnamefont
  {V.}~\bibnamefont {Khemani}}, \bibinfo {author} {\bibfnamefont {D.~A.}\
  \bibnamefont {Huse}}, \bibinfo {author} {\bibfnamefont {I.}~\bibnamefont
  {Bloch}},\ and\ \bibinfo {author} {\bibfnamefont {C.}~\bibnamefont {Gross}},\
  }\bibfield  {title} {\bibinfo {title} {{Exploring the many-body localization
  transition in two dimensions}},\ }\href@noop {} {\bibfield  {journal}
  {\bibinfo  {journal} {Science}\ }\textbf {\bibinfo {volume} {352}},\ \bibinfo
  {pages} {1547} (\bibinfo {year} {2016})}\BibitemShut {NoStop}%
\bibitem [{\citenamefont {Nandkishore}\ and\ \citenamefont
  {Huse}(2015)}]{Huse_review_2015}%
  \BibitemOpen
  \bibfield  {author} {\bibinfo {author} {\bibfnamefont {R.}~\bibnamefont
  {Nandkishore}}\ and\ \bibinfo {author} {\bibfnamefont {D.~A.}\ \bibnamefont
  {Huse}},\ }\bibfield  {title} {\bibinfo {title} {Many-body localization and
  thermalization in quantum statistical mechanics},\ }\href
  {https://doi.org/10.1146/annurev-conmatphys-031214-014726} {\bibfield
  {journal} {\bibinfo  {journal} {Annual Review of Condensed Matter Physics}\
  }\textbf {\bibinfo {volume} {6}},\ \bibinfo {pages} {15} (\bibinfo {year}
  {2015})},\ \Eprint
  {https://arxiv.org/abs/https://doi.org/10.1146/annurev-conmatphys-031214-014726}
  {https://doi.org/10.1146/annurev-conmatphys-031214-014726} \BibitemShut
  {NoStop}%
\bibitem [{\citenamefont {Altman}\ and\ \citenamefont
  {Vosk}(2015)}]{Altman_review_2015}%
  \BibitemOpen
  \bibfield  {author} {\bibinfo {author} {\bibfnamefont {E.}~\bibnamefont
  {Altman}}\ and\ \bibinfo {author} {\bibfnamefont {R.}~\bibnamefont {Vosk}},\
  }\bibfield  {title} {\bibinfo {title} {Universal dynamics and renormalization
  in many-body-localized systems},\ }\href
  {https://doi.org/10.1146/annurev-conmatphys-031214-014701} {\bibfield
  {journal} {\bibinfo  {journal} {Annual Review of Condensed Matter Physics}\
  }\textbf {\bibinfo {volume} {6}},\ \bibinfo {pages} {383} (\bibinfo {year}
  {2015})},\ \Eprint
  {https://arxiv.org/abs/https://doi.org/10.1146/annurev-conmatphys-031214-014701}
  {https://doi.org/10.1146/annurev-conmatphys-031214-014701} \BibitemShut
  {NoStop}%
\bibitem [{\citenamefont {Abanin}\ \emph {et~al.}(2019)\citenamefont {Abanin},
  \citenamefont {Altman}, \citenamefont {Bloch},\ and\ \citenamefont
  {Serbyn}}]{Abanin_RevModPhys_2019}%
  \BibitemOpen
  \bibfield  {author} {\bibinfo {author} {\bibfnamefont {D.~A.}\ \bibnamefont
  {Abanin}}, \bibinfo {author} {\bibfnamefont {E.}~\bibnamefont {Altman}},
  \bibinfo {author} {\bibfnamefont {I.}~\bibnamefont {Bloch}},\ and\ \bibinfo
  {author} {\bibfnamefont {M.}~\bibnamefont {Serbyn}},\ }\bibfield  {title}
  {\bibinfo {title} {Colloquium: Many-body localization, thermalization, and
  entanglement},\ }\href {https://doi.org/10.1103/RevModPhys.91.021001}
  {\bibfield  {journal} {\bibinfo  {journal} {Rev. Mod. Phys.}\ }\textbf
  {\bibinfo {volume} {91}},\ \bibinfo {pages} {021001} (\bibinfo {year}
  {2019})}\BibitemShut {NoStop}%
\bibitem [{\citenamefont {Alet}\ and\ \citenamefont
  {Laflorencie}(2018)}]{ALET_review_2017}%
  \BibitemOpen
  \bibfield  {author} {\bibinfo {author} {\bibfnamefont {F.}~\bibnamefont
  {Alet}}\ and\ \bibinfo {author} {\bibfnamefont {N.}~\bibnamefont
  {Laflorencie}},\ }\bibfield  {title} {\bibinfo {title} {Many-body
  localization: An introduction and selected topics},\ }\href
  {https://doi.org/https://doi.org/10.1016/j.crhy.2018.03.003} {\bibfield
  {journal} {\bibinfo  {journal} {Comptes Rendus Physique}\ }\textbf {\bibinfo
  {volume} {19}},\ \bibinfo {pages} {498} (\bibinfo {year} {2018})},\ \bibinfo
  {note} {quantum simulation / Simulation quantique}\BibitemShut {NoStop}%
\bibitem [{\citenamefont {Imbrie}\ \emph {et~al.}(2017)\citenamefont {Imbrie},
  \citenamefont {Ros},\ and\ \citenamefont {Scardicchio}}]{Imbrie_2017}%
  \BibitemOpen
  \bibfield  {author} {\bibinfo {author} {\bibfnamefont {J.~Z.}\ \bibnamefont
  {Imbrie}}, \bibinfo {author} {\bibfnamefont {V.}~\bibnamefont {Ros}},\ and\
  \bibinfo {author} {\bibfnamefont {A.}~\bibnamefont {Scardicchio}},\
  }\bibfield  {title} {\bibinfo {title} {Local integrals of motion in many-body
  localized systems},\ }\href
  {https://doi.org/https://doi.org/10.1002/andp.201600278} {\bibfield
  {journal} {\bibinfo  {journal} {Annalen der Physik}\ }\textbf {\bibinfo
  {volume} {529}},\ \bibinfo {pages} {1600278} (\bibinfo {year}
  {2017})}\BibitemShut {NoStop}%
\bibitem [{\citenamefont {De~Tomasi}\ \emph {et~al.}(2019)\citenamefont
  {De~Tomasi}, \citenamefont {Pollmann},\ and\ \citenamefont
  {Heyl}}]{DeTomasi2019Efficient}%
  \BibitemOpen
  \bibfield  {author} {\bibinfo {author} {\bibfnamefont {G.}~\bibnamefont
  {De~Tomasi}}, \bibinfo {author} {\bibfnamefont {F.}~\bibnamefont
  {Pollmann}},\ and\ \bibinfo {author} {\bibfnamefont {M.}~\bibnamefont
  {Heyl}},\ }\bibfield  {title} {\bibinfo {title} {Efficiently solving the
  dynamics of many-body localized systems at strong disorder},\ }\href
  {https://doi.org/10.1103/PhysRevB.99.241114} {\bibfield  {journal} {\bibinfo
  {journal} {Phys. Rev. B}\ }\textbf {\bibinfo {volume} {99}},\ \bibinfo
  {pages} {241114} (\bibinfo {year} {2019})}\BibitemShut {NoStop}%
\bibitem [{\citenamefont {Huse}\ \emph {et~al.}(2014)\citenamefont {Huse},
  \citenamefont {Nandkishore},\ and\ \citenamefont
  {Oganesyan}}]{Huse_2014_fully}%
  \BibitemOpen
  \bibfield  {author} {\bibinfo {author} {\bibfnamefont {D.~A.}\ \bibnamefont
  {Huse}}, \bibinfo {author} {\bibfnamefont {R.}~\bibnamefont {Nandkishore}},\
  and\ \bibinfo {author} {\bibfnamefont {V.}~\bibnamefont {Oganesyan}},\
  }\bibfield  {title} {\bibinfo {title} {Phenomenology of fully
  many-body-localized systems},\ }\href
  {https://doi.org/10.1103/PhysRevB.90.174202} {\bibfield  {journal} {\bibinfo
  {journal} {Phys. Rev. B}\ }\textbf {\bibinfo {volume} {90}},\ \bibinfo
  {pages} {174202} (\bibinfo {year} {2014})}\BibitemShut {NoStop}%
\bibitem [{\citenamefont {Serbyn}\ \emph
  {et~al.}(2013{\natexlab{a}})\citenamefont {Serbyn}, \citenamefont
  {Papi\ifmmode~\acute{c}\else \'{c}\fi{}},\ and\ \citenamefont
  {Abanin}}]{Local_Serbyn_2013}%
  \BibitemOpen
  \bibfield  {author} {\bibinfo {author} {\bibfnamefont {M.}~\bibnamefont
  {Serbyn}}, \bibinfo {author} {\bibfnamefont {Z.}~\bibnamefont
  {Papi\ifmmode~\acute{c}\else \'{c}\fi{}}},\ and\ \bibinfo {author}
  {\bibfnamefont {D.~A.}\ \bibnamefont {Abanin}},\ }\bibfield  {title}
  {\bibinfo {title} {Local conservation laws and the structure of the many-body
  localized states},\ }\href {https://doi.org/10.1103/PhysRevLett.111.127201}
  {\bibfield  {journal} {\bibinfo  {journal} {Phys. Rev. Lett.}\ }\textbf
  {\bibinfo {volume} {111}},\ \bibinfo {pages} {127201} (\bibinfo {year}
  {2013}{\natexlab{a}})}\BibitemShut {NoStop}%
\bibitem [{\citenamefont {Chandran}\ \emph {et~al.}(2015)\citenamefont
  {Chandran}, \citenamefont {Kim}, \citenamefont {Vidal},\ and\ \citenamefont
  {Abanin}}]{Chandran_2015}%
  \BibitemOpen
  \bibfield  {author} {\bibinfo {author} {\bibfnamefont {A.}~\bibnamefont
  {Chandran}}, \bibinfo {author} {\bibfnamefont {I.~H.}\ \bibnamefont {Kim}},
  \bibinfo {author} {\bibfnamefont {G.}~\bibnamefont {Vidal}},\ and\ \bibinfo
  {author} {\bibfnamefont {D.~A.}\ \bibnamefont {Abanin}},\ }\bibfield  {title}
  {\bibinfo {title} {Constructing local integrals of motion in the many-body
  localized phase},\ }\href {https://doi.org/10.1103/PhysRevB.91.085425}
  {\bibfield  {journal} {\bibinfo  {journal} {Phys. Rev. B}\ }\textbf {\bibinfo
  {volume} {91}},\ \bibinfo {pages} {085425} (\bibinfo {year}
  {2015})}\BibitemShut {NoStop}%
\bibitem [{\citenamefont {Bardarson}\ \emph {et~al.}(2012)\citenamefont
  {Bardarson}, \citenamefont {Pollmann},\ and\ \citenamefont
  {Moore}}]{Bardarson2012unbounded}%
  \BibitemOpen
  \bibfield  {author} {\bibinfo {author} {\bibfnamefont {J.~H.}\ \bibnamefont
  {Bardarson}}, \bibinfo {author} {\bibfnamefont {F.}~\bibnamefont
  {Pollmann}},\ and\ \bibinfo {author} {\bibfnamefont {J.~E.}\ \bibnamefont
  {Moore}},\ }\bibfield  {title} {\bibinfo {title} {{Unbounded Growth of
  Entanglement in Models of Many-Body Localization}},\ }\href
  {https://journals.aps.org/prl/pdf/10.1103/PhysRevLett.109.017202} {\bibfield
  {journal} {\bibinfo  {journal} {Phys. Rev. Lett.}\ }\textbf {\bibinfo
  {volume} {109}},\ \bibinfo {pages} {017202} (\bibinfo {year}
  {2012})}\BibitemShut {NoStop}%
\bibitem [{\citenamefont {Serbyn}\ \emph
  {et~al.}(2013{\natexlab{b}})\citenamefont {Serbyn}, \citenamefont
  {Papi{\'c}},\ and\ \citenamefont {Abanin}}]{Serbyn2013universal}%
  \BibitemOpen
  \bibfield  {author} {\bibinfo {author} {\bibfnamefont {M.}~\bibnamefont
  {Serbyn}}, \bibinfo {author} {\bibfnamefont {Z.}~\bibnamefont {Papi{\'c}}},\
  and\ \bibinfo {author} {\bibfnamefont {D.~A.}\ \bibnamefont {Abanin}},\
  }\bibfield  {title} {\bibinfo {title} {{Universal slow growth of entanglement
  in interacting strongly disordered systems}},\ }\href
  {https://doi.org/10.1103/PhysRevLett.110.260601} {\bibfield  {journal}
  {\bibinfo  {journal} {Phys. Rev. Lett.}\ }\textbf {\bibinfo {volume} {110}},\
  \bibinfo {pages} {260601} (\bibinfo {year} {2013}{\natexlab{b}})}\BibitemShut
  {NoStop}%
\bibitem [{\citenamefont {\ifmmode \check{Z}\else
  \v{Z}\fi{}nidari\ifmmode~\check{c}\else \v{c}\fi{}}\ \emph
  {et~al.}(2008)\citenamefont {\ifmmode \check{Z}\else
  \v{Z}\fi{}nidari\ifmmode~\check{c}\else \v{c}\fi{}}, \citenamefont {Prosen},\
  and\ \citenamefont {Prelov\ifmmode~\check{s}\else
  \v{s}\fi{}ek}}]{Prosen_2008}%
  \BibitemOpen
  \bibfield  {author} {\bibinfo {author} {\bibfnamefont {M.}~\bibnamefont
  {\ifmmode \check{Z}\else \v{Z}\fi{}nidari\ifmmode~\check{c}\else
  \v{c}\fi{}}}, \bibinfo {author} {\bibfnamefont {T.~c.~v.}\ \bibnamefont
  {Prosen}},\ and\ \bibinfo {author} {\bibfnamefont {P.}~\bibnamefont
  {Prelov\ifmmode~\check{s}\else \v{s}\fi{}ek}},\ }\bibfield  {title} {\bibinfo
  {title} {Many-body localization in the heisenberg $xxz$ magnet in a random
  field},\ }\href {https://doi.org/10.1103/PhysRevB.77.064426} {\bibfield
  {journal} {\bibinfo  {journal} {Phys. Rev. B}\ }\textbf {\bibinfo {volume}
  {77}},\ \bibinfo {pages} {064426} (\bibinfo {year} {2008})}\BibitemShut
  {NoStop}%
\bibitem [{\citenamefont {De~Tomasi}\ \emph
  {et~al.}(2017{\natexlab{a}})\citenamefont {De~Tomasi}, \citenamefont {Bera},
  \citenamefont {Bardarson},\ and\ \citenamefont {Pollmann}}]{DeTomasi2017QMI}%
  \BibitemOpen
  \bibfield  {author} {\bibinfo {author} {\bibfnamefont {G.}~\bibnamefont
  {De~Tomasi}}, \bibinfo {author} {\bibfnamefont {S.}~\bibnamefont {Bera}},
  \bibinfo {author} {\bibfnamefont {J.~H.}\ \bibnamefont {Bardarson}},\ and\
  \bibinfo {author} {\bibfnamefont {F.}~\bibnamefont {Pollmann}},\ }\bibfield
  {title} {\bibinfo {title} {Quantum mutual information as a probe for
  many-body localization},\ }\href
  {https://doi.org/10.1103/PhysRevLett.118.016804} {\bibfield  {journal}
  {\bibinfo  {journal} {Phys. Rev. Lett.}\ }\textbf {\bibinfo {volume} {118}},\
  \bibinfo {pages} {016804} (\bibinfo {year} {2017}{\natexlab{a}})}\BibitemShut
  {NoStop}%
\bibitem [{\citenamefont {Serbyn}\ \emph
  {et~al.}(2014{\natexlab{a}})\citenamefont {Serbyn}, \citenamefont
  {Papi{\'c}},\ and\ \citenamefont {Abanin}}]{Serbyn2014quenches}%
  \BibitemOpen
  \bibfield  {author} {\bibinfo {author} {\bibfnamefont {M.}~\bibnamefont
  {Serbyn}}, \bibinfo {author} {\bibfnamefont {Z.}~\bibnamefont {Papi{\'c}}},\
  and\ \bibinfo {author} {\bibfnamefont {D.~A.}\ \bibnamefont {Abanin}},\
  }\bibfield  {title} {\bibinfo {title} {{Quantum quenches in the many-body
  localized phase}},\ }\href@noop {} {\bibfield  {journal} {\bibinfo  {journal}
  {Phys. Rev. B}\ }\textbf {\bibinfo {volume} {90}},\ \bibinfo {pages} {174302}
  (\bibinfo {year} {2014}{\natexlab{a}})}\BibitemShut {NoStop}%
\bibitem [{\citenamefont {Roy}\ \emph {et~al.}(2015)\citenamefont {Roy},
  \citenamefont {Singh},\ and\ \citenamefont {Moessner}}]{Rajeev_2015}%
  \BibitemOpen
  \bibfield  {author} {\bibinfo {author} {\bibfnamefont {D.}~\bibnamefont
  {Roy}}, \bibinfo {author} {\bibfnamefont {R.}~\bibnamefont {Singh}},\ and\
  \bibinfo {author} {\bibfnamefont {R.}~\bibnamefont {Moessner}},\ }\bibfield
  {title} {\bibinfo {title} {Probing many-body localization by spin noise
  spectroscopy},\ }\href {https://doi.org/10.1103/PhysRevB.92.180205}
  {\bibfield  {journal} {\bibinfo  {journal} {Phys. Rev. B}\ }\textbf {\bibinfo
  {volume} {92}},\ \bibinfo {pages} {180205} (\bibinfo {year}
  {2015})}\BibitemShut {NoStop}%
\bibitem [{\citenamefont {Serbyn}\ \emph
  {et~al.}(2014{\natexlab{b}})\citenamefont {Serbyn}, \citenamefont {Knap},
  \citenamefont {Gopalakrishnan}, \citenamefont {Papi\ifmmode~\acute{c}\else
  \'{c}\fi{}}, \citenamefont {Yao}, \citenamefont {Laumann}, \citenamefont
  {Abanin}, \citenamefont {Lukin},\ and\ \citenamefont
  {Demler}}]{Serbyn_inter_2014}%
  \BibitemOpen
  \bibfield  {author} {\bibinfo {author} {\bibfnamefont {M.}~\bibnamefont
  {Serbyn}}, \bibinfo {author} {\bibfnamefont {M.}~\bibnamefont {Knap}},
  \bibinfo {author} {\bibfnamefont {S.}~\bibnamefont {Gopalakrishnan}},
  \bibinfo {author} {\bibfnamefont {Z.}~\bibnamefont
  {Papi\ifmmode~\acute{c}\else \'{c}\fi{}}}, \bibinfo {author} {\bibfnamefont
  {N.~Y.}\ \bibnamefont {Yao}}, \bibinfo {author} {\bibfnamefont {C.~R.}\
  \bibnamefont {Laumann}}, \bibinfo {author} {\bibfnamefont {D.~A.}\
  \bibnamefont {Abanin}}, \bibinfo {author} {\bibfnamefont {M.~D.}\
  \bibnamefont {Lukin}},\ and\ \bibinfo {author} {\bibfnamefont {E.~A.}\
  \bibnamefont {Demler}},\ }\bibfield  {title} {\bibinfo {title}
  {Interferometric probes of many-body localization},\ }\href
  {https://doi.org/10.1103/PhysRevLett.113.147204} {\bibfield  {journal}
  {\bibinfo  {journal} {Phys. Rev. Lett.}\ }\textbf {\bibinfo {volume} {113}},\
  \bibinfo {pages} {147204} (\bibinfo {year} {2014}{\natexlab{b}})}\BibitemShut
  {NoStop}%
\bibitem [{\citenamefont {van Nieuwenburg}\ \emph {et~al.}(2017)\citenamefont
  {van Nieuwenburg}, \citenamefont {Liu},\ and\ \citenamefont
  {Huber}}]{Nieuwenburg2017Confusion}%
  \BibitemOpen
  \bibfield  {author} {\bibinfo {author} {\bibfnamefont {E.~P.~L.}\
  \bibnamefont {van Nieuwenburg}}, \bibinfo {author} {\bibfnamefont {Y.-H.}\
  \bibnamefont {Liu}},\ and\ \bibinfo {author} {\bibfnamefont {S.~D.}\
  \bibnamefont {Huber}},\ }\bibfield  {title} {\bibinfo {title} {Learning phase
  transitions by confusion},\ }\href {https://doi.org/10.1038/nphys4037}
  {\bibfield  {journal} {\bibinfo  {journal} {Nature Physics}\ }\textbf
  {\bibinfo {volume} {13}},\ \bibinfo {pages} {435–439} (\bibinfo {year}
  {2017})}\BibitemShut {NoStop}%
\bibitem [{\citenamefont {van Nieuwenburg}\ \emph {et~al.}(2019)\citenamefont
  {van Nieuwenburg}, \citenamefont {Baum},\ and\ \citenamefont
  {Refael}}]{Nieuwenburg2019Bloch}%
  \BibitemOpen
  \bibfield  {author} {\bibinfo {author} {\bibfnamefont {E.}~\bibnamefont {van
  Nieuwenburg}}, \bibinfo {author} {\bibfnamefont {Y.}~\bibnamefont {Baum}},\
  and\ \bibinfo {author} {\bibfnamefont {G.}~\bibnamefont {Refael}},\
  }\bibfield  {title} {\bibinfo {title} {From bloch oscillations to many-body
  localization in clean interacting systems},\ }\href
  {https://doi.org/10.1073/pnas.1819316116} {\bibfield  {journal} {\bibinfo
  {journal} {Proceedings of the National Academy of Sciences}\ }\textbf
  {\bibinfo {volume} {116}},\ \bibinfo {pages} {9269–9274} (\bibinfo {year}
  {2019})}\BibitemShut {NoStop}%
\bibitem [{\citenamefont {Carrasquilla}\ and\ \citenamefont
  {Melko}(2017)}]{Carrasquilla2017MLPhase}%
  \BibitemOpen
  \bibfield  {author} {\bibinfo {author} {\bibfnamefont {J.}~\bibnamefont
  {Carrasquilla}}\ and\ \bibinfo {author} {\bibfnamefont {R.~G.}\ \bibnamefont
  {Melko}},\ }\bibfield  {title} {\bibinfo {title} {Machine learning phases of
  matter},\ }\href {https://doi.org/10.1038/nphys4035} {\bibfield  {journal}
  {\bibinfo  {journal} {Nature Physics}\ }\textbf {\bibinfo {volume} {13}},\
  \bibinfo {pages} {431–434} (\bibinfo {year} {2017})}\BibitemShut {NoStop}%
\bibitem [{\citenamefont {Wetzel}(2017)}]{Wetzel2017Unsupervised}%
  \BibitemOpen
  \bibfield  {author} {\bibinfo {author} {\bibfnamefont {S.~J.}\ \bibnamefont
  {Wetzel}},\ }\bibfield  {title} {\bibinfo {title} {Unsupervised learning of
  phase transitions: From principal component analysis to variational
  autoencoders},\ }\bibfield  {journal} {\bibinfo  {journal} {Physical Review
  E}\ }\textbf {\bibinfo {volume} {96}},\ \href
  {https://doi.org/10.1103/physreve.96.022140} {10.1103/physreve.96.022140}
  (\bibinfo {year} {2017})\BibitemShut {NoStop}%
\bibitem [{\citenamefont {Ch’ng}\ \emph {et~al.}(2017)\citenamefont
  {Ch’ng}, \citenamefont {Carrasquilla}, \citenamefont {Melko},\ and\
  \citenamefont {Khatami}}]{Chng2017Correlated}%
  \BibitemOpen
  \bibfield  {author} {\bibinfo {author} {\bibfnamefont {K.}~\bibnamefont
  {Ch’ng}}, \bibinfo {author} {\bibfnamefont {J.}~\bibnamefont
  {Carrasquilla}}, \bibinfo {author} {\bibfnamefont {R.~G.}\ \bibnamefont
  {Melko}},\ and\ \bibinfo {author} {\bibfnamefont {E.}~\bibnamefont
  {Khatami}},\ }\bibfield  {title} {\bibinfo {title} {Machine learning phases
  of strongly correlated fermions},\ }\bibfield  {journal} {\bibinfo  {journal}
  {Physical Review X}\ }\textbf {\bibinfo {volume} {7}},\ \href
  {https://doi.org/10.1103/physrevx.7.031038} {10.1103/physrevx.7.031038}
  (\bibinfo {year} {2017})\BibitemShut {NoStop}%
\bibitem [{\citenamefont {Carleo}\ and\ \citenamefont
  {Troyer}(2017)}]{Carleo2017Solving}%
  \BibitemOpen
  \bibfield  {author} {\bibinfo {author} {\bibfnamefont {G.}~\bibnamefont
  {Carleo}}\ and\ \bibinfo {author} {\bibfnamefont {M.}~\bibnamefont
  {Troyer}},\ }\bibfield  {title} {\bibinfo {title} {Solving the quantum
  many-body problem with artificial neural networks},\ }\href
  {https://doi.org/10.1126/science.aag2302} {\bibfield  {journal} {\bibinfo
  {journal} {Science}\ }\textbf {\bibinfo {volume} {355}},\ \bibinfo {pages}
  {602–606} (\bibinfo {year} {2017})}\BibitemShut {NoStop}%
\bibitem [{\citenamefont {Zhang}\ \emph {et~al.}(2018)\citenamefont {Zhang},
  \citenamefont {Shen},\ and\ \citenamefont {Zhai}}]{Zhang2018Topology}%
  \BibitemOpen
  \bibfield  {author} {\bibinfo {author} {\bibfnamefont {P.}~\bibnamefont
  {Zhang}}, \bibinfo {author} {\bibfnamefont {H.}~\bibnamefont {Shen}},\ and\
  \bibinfo {author} {\bibfnamefont {H.}~\bibnamefont {Zhai}},\ }\bibfield
  {title} {\bibinfo {title} {Machine learning topological invariants with
  neural networks},\ }\bibfield  {journal} {\bibinfo  {journal} {Physical
  Review Letters}\ }\textbf {\bibinfo {volume} {120}},\ \href
  {https://doi.org/10.1103/physrevlett.120.066401}
  {10.1103/physrevlett.120.066401} (\bibinfo {year} {2018})\BibitemShut
  {NoStop}%
\bibitem [{\citenamefont {{Bohrdt}}\ \emph {et~al.}(2019)\citenamefont
  {{Bohrdt}}, \citenamefont {{Chiu}}, \citenamefont {{Ji}}, \citenamefont
  {{Xu}}, \citenamefont {{Greif}}, \citenamefont {{Greiner}}, \citenamefont
  {{Demler}}, \citenamefont {{Grusdt}},\ and\ \citenamefont
  {{Knap}}}]{Bohrdt2019Snapshots}%
  \BibitemOpen
  \bibfield  {author} {\bibinfo {author} {\bibfnamefont {A.}~\bibnamefont
  {{Bohrdt}}}, \bibinfo {author} {\bibfnamefont {C.~S.}\ \bibnamefont
  {{Chiu}}}, \bibinfo {author} {\bibfnamefont {G.}~\bibnamefont {{Ji}}},
  \bibinfo {author} {\bibfnamefont {M.}~\bibnamefont {{Xu}}}, \bibinfo {author}
  {\bibfnamefont {D.}~\bibnamefont {{Greif}}}, \bibinfo {author} {\bibfnamefont
  {M.}~\bibnamefont {{Greiner}}}, \bibinfo {author} {\bibfnamefont
  {E.}~\bibnamefont {{Demler}}}, \bibinfo {author} {\bibfnamefont
  {F.}~\bibnamefont {{Grusdt}}},\ and\ \bibinfo {author} {\bibfnamefont
  {M.}~\bibnamefont {{Knap}}},\ }\bibfield  {title} {\bibinfo {title}
  {{Classifying snapshots of the doped Hubbard model with machine learning}},\
  }\href {https://doi.org/10.1038/s41567-019-0565-x} {\bibfield  {journal}
  {\bibinfo  {journal} {Nature Physics}\ }\textbf {\bibinfo {volume} {15}},\
  \bibinfo {pages} {921} (\bibinfo {year} {2019})},\ \Eprint
  {https://arxiv.org/abs/1811.12425} {arXiv:1811.12425 [cond-mat.quant-gas]}
  \BibitemShut {NoStop}%
\bibitem [{\citenamefont {Scheurer}\ and\ \citenamefont
  {Slager}(2020)}]{Robert_2020}%
  \BibitemOpen
  \bibfield  {author} {\bibinfo {author} {\bibfnamefont {M.~S.}\ \bibnamefont
  {Scheurer}}\ and\ \bibinfo {author} {\bibfnamefont {R.-J.}\ \bibnamefont
  {Slager}},\ }\bibfield  {title} {\bibinfo {title} {Unsupervised machine
  learning and band topology},\ }\href
  {https://doi.org/10.1103/PhysRevLett.124.226401} {\bibfield  {journal}
  {\bibinfo  {journal} {Phys. Rev. Lett.}\ }\textbf {\bibinfo {volume} {124}},\
  \bibinfo {pages} {226401} (\bibinfo {year} {2020})}\BibitemShut {NoStop}%
\bibitem [{\citenamefont {{Neupert}}\ \emph {et~al.}(2021)\citenamefont
  {{Neupert}}, \citenamefont {{Fischer}}, \citenamefont {{Greplova}},
  \citenamefont {{Choo}},\ and\ \citenamefont {{Denner}}}]{Titus_2020}%
  \BibitemOpen
  \bibfield  {author} {\bibinfo {author} {\bibfnamefont {T.}~\bibnamefont
  {{Neupert}}}, \bibinfo {author} {\bibfnamefont {M.~H.}\ \bibnamefont
  {{Fischer}}}, \bibinfo {author} {\bibfnamefont {E.}~\bibnamefont
  {{Greplova}}}, \bibinfo {author} {\bibfnamefont {K.}~\bibnamefont {{Choo}}},\
  and\ \bibinfo {author} {\bibfnamefont {M.}~\bibnamefont {{Denner}}},\
  }\bibfield  {title} {\bibinfo {title} {{Introduction to Machine Learning for
  the Sciences}},\ }\href@noop {} {\bibfield  {journal} {\bibinfo  {journal}
  {arXiv e-prints}\ ,\ \bibinfo {eid} {arXiv:2102.04883}} (\bibinfo {year}
  {2021})},\ \Eprint {https://arxiv.org/abs/2102.04883} {arXiv:2102.04883
  [physics.comp-ph]} \BibitemShut {NoStop}%
\bibitem [{\citenamefont {Doggen}\ \emph {et~al.}(2018)\citenamefont {Doggen},
  \citenamefont {Schindler}, \citenamefont {Tikhonov}, \citenamefont {Mirlin},
  \citenamefont {Neupert}, \citenamefont {Polyakov},\ and\ \citenamefont
  {Gornyi}}]{Doggen2018LargeChains}%
  \BibitemOpen
  \bibfield  {author} {\bibinfo {author} {\bibfnamefont {E.~V.~H.}\
  \bibnamefont {Doggen}}, \bibinfo {author} {\bibfnamefont {F.}~\bibnamefont
  {Schindler}}, \bibinfo {author} {\bibfnamefont {K.~S.}\ \bibnamefont
  {Tikhonov}}, \bibinfo {author} {\bibfnamefont {A.~D.}\ \bibnamefont
  {Mirlin}}, \bibinfo {author} {\bibfnamefont {T.}~\bibnamefont {Neupert}},
  \bibinfo {author} {\bibfnamefont {D.~G.}\ \bibnamefont {Polyakov}},\ and\
  \bibinfo {author} {\bibfnamefont {I.~V.}\ \bibnamefont {Gornyi}},\ }\bibfield
   {title} {\bibinfo {title} {Many-body localization and delocalization in
  large quantum chains},\ }\href {https://doi.org/10.1103/PhysRevB.98.174202}
  {\bibfield  {journal} {\bibinfo  {journal} {Phys. Rev. B}\ }\textbf {\bibinfo
  {volume} {98}},\ \bibinfo {pages} {174202} (\bibinfo {year}
  {2018})}\BibitemShut {NoStop}%
\bibitem [{\citenamefont {{Hsu}}\ \emph {et~al.}(2018)\citenamefont {{Hsu}},
  \citenamefont {{Li}}, \citenamefont {{Deng}},\ and\ \citenamefont {{Das
  Sarma}}}]{Hsu2018Elusive}%
  \BibitemOpen
  \bibfield  {author} {\bibinfo {author} {\bibfnamefont {Y.-T.}\ \bibnamefont
  {{Hsu}}}, \bibinfo {author} {\bibfnamefont {X.}~\bibnamefont {{Li}}},
  \bibinfo {author} {\bibfnamefont {D.-L.}\ \bibnamefont {{Deng}}},\ and\
  \bibinfo {author} {\bibfnamefont {S.}~\bibnamefont {{Das Sarma}}},\
  }\bibfield  {title} {\bibinfo {title} {{Machine Learning Many-Body
  Localization: Search for the Elusive Nonergodic Metal}},\ }\href
  {https://doi.org/10.1103/PhysRevLett.121.245701} {\bibfield  {journal}
  {\bibinfo  {journal} {\prl}\ }\textbf {\bibinfo {volume} {121}},\ \bibinfo
  {eid} {245701} (\bibinfo {year} {2018})},\ \Eprint
  {https://arxiv.org/abs/1805.12138} {arXiv:1805.12138 [cond-mat.dis-nn]}
  \BibitemShut {NoStop}%
\bibitem [{\citenamefont {Huembeli}\ \emph {et~al.}(2019)\citenamefont
  {Huembeli}, \citenamefont {Dauphin}, \citenamefont {Wittek},\ and\
  \citenamefont {Gogolin}}]{Huembeli2019Automated}%
  \BibitemOpen
  \bibfield  {author} {\bibinfo {author} {\bibfnamefont {P.}~\bibnamefont
  {Huembeli}}, \bibinfo {author} {\bibfnamefont {A.}~\bibnamefont {Dauphin}},
  \bibinfo {author} {\bibfnamefont {P.}~\bibnamefont {Wittek}},\ and\ \bibinfo
  {author} {\bibfnamefont {C.}~\bibnamefont {Gogolin}},\ }\bibfield  {title}
  {\bibinfo {title} {Automated discovery of characteristic features of phase
  transitions in many-body localization},\ }\href
  {https://doi.org/10.1103/PhysRevB.99.104106} {\bibfield  {journal} {\bibinfo
  {journal} {Phys. Rev. B}\ }\textbf {\bibinfo {volume} {99}},\ \bibinfo
  {pages} {104106} (\bibinfo {year} {2019})}\BibitemShut {NoStop}%
\bibitem [{\citenamefont {Rao}(2018)}]{Rao2018Random}%
  \BibitemOpen
  \bibfield  {author} {\bibinfo {author} {\bibfnamefont {W.-J.}\ \bibnamefont
  {Rao}},\ }\bibfield  {title} {\bibinfo {title} {Machine learning the
  many-body localization transition in random spin systems},\ }\href
  {https://doi.org/10.1088/1361-648x/aaddc6} {\bibfield  {journal} {\bibinfo
  {journal} {Journal of Physics: Condensed Matter}\ }\textbf {\bibinfo {volume}
  {30}},\ \bibinfo {pages} {395902} (\bibinfo {year} {2018})}\BibitemShut
  {NoStop}%
\bibitem [{\citenamefont {Schindler}\ \emph {et~al.}(2017)\citenamefont
  {Schindler}, \citenamefont {Regnault},\ and\ \citenamefont
  {Neupert}}]{Schindler2017Probing}%
  \BibitemOpen
  \bibfield  {author} {\bibinfo {author} {\bibfnamefont {F.}~\bibnamefont
  {Schindler}}, \bibinfo {author} {\bibfnamefont {N.}~\bibnamefont
  {Regnault}},\ and\ \bibinfo {author} {\bibfnamefont {T.}~\bibnamefont
  {Neupert}},\ }\bibfield  {title} {\bibinfo {title} {Probing many-body
  localization with neural networks},\ }\bibfield  {journal} {\bibinfo
  {journal} {Physical Review B}\ }\textbf {\bibinfo {volume} {95}},\ \href
  {https://doi.org/10.1103/physrevb.95.245134} {10.1103/physrevb.95.245134}
  (\bibinfo {year} {2017})\BibitemShut {NoStop}%
\bibitem [{\citenamefont {Kausar}\ \emph {et~al.}(2020)\citenamefont {Kausar},
  \citenamefont {Rao},\ and\ \citenamefont {Wan}}]{Kausar2020Learning}%
  \BibitemOpen
  \bibfield  {author} {\bibinfo {author} {\bibfnamefont {R.}~\bibnamefont
  {Kausar}}, \bibinfo {author} {\bibfnamefont {W.-J.}\ \bibnamefont {Rao}},\
  and\ \bibinfo {author} {\bibfnamefont {X.}~\bibnamefont {Wan}},\ }\bibfield
  {title} {\bibinfo {title} {Learning what a machine learns in a many-body
  localization transition},\ }\href {https://doi.org/10.1088/1361-648x/ab9f09}
  {\bibfield  {journal} {\bibinfo  {journal} {Journal of Physics: Condensed
  Matter}\ }\textbf {\bibinfo {volume} {32}},\ \bibinfo {pages} {415605}
  (\bibinfo {year} {2020})}\BibitemShut {NoStop}%
\bibitem [{\citenamefont {{Th{\'e}veniaut}}\ and\ \citenamefont
  {{Alet}}(2019)}]{Theveniaut2019Precise}%
  \BibitemOpen
  \bibfield  {author} {\bibinfo {author} {\bibfnamefont {H.}~\bibnamefont
  {{Th{\'e}veniaut}}}\ and\ \bibinfo {author} {\bibfnamefont {F.}~\bibnamefont
  {{Alet}}},\ }\bibfield  {title} {\bibinfo {title} {{Neural network setups for
  a precise detection of the many-body localization transition: Finite-size
  scaling and limitations}},\ }\href
  {https://doi.org/10.1103/PhysRevB.100.224202} {\bibfield  {journal} {\bibinfo
   {journal} {\prb}\ }\textbf {\bibinfo {volume} {100}},\ \bibinfo {eid}
  {224202} (\bibinfo {year} {2019})},\ \Eprint
  {https://arxiv.org/abs/1904.13165} {arXiv:1904.13165 [cond-mat.dis-nn]}
  \BibitemShut {NoStop}%
\bibitem [{\citenamefont {Zhang}\ \emph {et~al.}(2019)\citenamefont {Zhang},
  \citenamefont {Wang},\ and\ \citenamefont {Wang}}]{Zhang2019Interpretable}%
  \BibitemOpen
  \bibfield  {author} {\bibinfo {author} {\bibfnamefont {W.}~\bibnamefont
  {Zhang}}, \bibinfo {author} {\bibfnamefont {L.}~\bibnamefont {Wang}},\ and\
  \bibinfo {author} {\bibfnamefont {Z.}~\bibnamefont {Wang}},\ }\bibfield
  {title} {\bibinfo {title} {Interpretable machine learning study of the
  many-body localization transition in disordered quantum ising spin chains},\
  }\bibfield  {journal} {\bibinfo  {journal} {Physical Review B}\ }\textbf
  {\bibinfo {volume} {99}},\ \href {https://doi.org/10.1103/physrevb.99.054208}
  {10.1103/physrevb.99.054208} (\bibinfo {year} {2019})\BibitemShut {NoStop}%
\bibitem [{\citenamefont {van Nieuwenburg}\ \emph {et~al.}(2018)\citenamefont
  {van Nieuwenburg}, \citenamefont {Bairey},\ and\ \citenamefont
  {Refael}}]{Nieuwenburg2018PTDynamics}%
  \BibitemOpen
  \bibfield  {author} {\bibinfo {author} {\bibfnamefont {E.}~\bibnamefont {van
  Nieuwenburg}}, \bibinfo {author} {\bibfnamefont {E.}~\bibnamefont {Bairey}},\
  and\ \bibinfo {author} {\bibfnamefont {G.}~\bibnamefont {Refael}},\
  }\bibfield  {title} {\bibinfo {title} {Learning phase transitions from
  dynamics},\ }\href {https://doi.org/10.1103/PhysRevB.98.060301} {\bibfield
  {journal} {\bibinfo  {journal} {Phys. Rev. B}\ }\textbf {\bibinfo {volume}
  {98}},\ \bibinfo {pages} {060301} (\bibinfo {year} {2018})}\BibitemShut
  {NoStop}%
\bibitem [{\citenamefont {{Bohrdt}}\ \emph {et~al.}(2020)\citenamefont
  {{Bohrdt}}, \citenamefont {{Kim}}, \citenamefont {{Lukin}}, \citenamefont
  {{Rispoli}}, \citenamefont {{Schittko}}, \citenamefont {{Knap}},
  \citenamefont {{Greiner}},\ and\ \citenamefont
  {{L{\'e}onard}}}]{Bohrdt_2020}%
  \BibitemOpen
  \bibfield  {author} {\bibinfo {author} {\bibfnamefont {A.}~\bibnamefont
  {{Bohrdt}}}, \bibinfo {author} {\bibfnamefont {S.}~\bibnamefont {{Kim}}},
  \bibinfo {author} {\bibfnamefont {A.}~\bibnamefont {{Lukin}}}, \bibinfo
  {author} {\bibfnamefont {M.}~\bibnamefont {{Rispoli}}}, \bibinfo {author}
  {\bibfnamefont {R.}~\bibnamefont {{Schittko}}}, \bibinfo {author}
  {\bibfnamefont {M.}~\bibnamefont {{Knap}}}, \bibinfo {author} {\bibfnamefont
  {M.}~\bibnamefont {{Greiner}}},\ and\ \bibinfo {author} {\bibfnamefont
  {J.}~\bibnamefont {{L{\'e}onard}}},\ }\bibfield  {title} {\bibinfo {title}
  {{Analyzing non-equilibrium quantum states through snapshots with artificial
  neural networks}},\ }\href@noop {} {\bibfield  {journal} {\bibinfo  {journal}
  {arXiv e-prints}\ ,\ \bibinfo {eid} {arXiv:2012.11586}} (\bibinfo {year}
  {2020})},\ \Eprint {https://arxiv.org/abs/2012.11586} {arXiv:2012.11586
  [cond-mat.quant-gas]} \BibitemShut {NoStop}%
\bibitem [{\citenamefont {Bakr}\ \emph {et~al.}(2009)\citenamefont {Bakr},
  \citenamefont {Gillen}, \citenamefont {Peng}, \citenamefont {F{\"o}lling},\
  and\ \citenamefont {Greiner}}]{Bakr2009}%
  \BibitemOpen
  \bibfield  {author} {\bibinfo {author} {\bibfnamefont {W.~S.}\ \bibnamefont
  {Bakr}}, \bibinfo {author} {\bibfnamefont {J.~I.}\ \bibnamefont {Gillen}},
  \bibinfo {author} {\bibfnamefont {A.}~\bibnamefont {Peng}}, \bibinfo {author}
  {\bibfnamefont {S.}~\bibnamefont {F{\"o}lling}},\ and\ \bibinfo {author}
  {\bibfnamefont {M.}~\bibnamefont {Greiner}},\ }\bibfield  {title} {\bibinfo
  {title} {A quantum gas microscope for detecting single atoms in a
  hubbard-regime optical lattice},\ }\href
  {https://doi.org/10.1038/nature08482} {\bibfield  {journal} {\bibinfo
  {journal} {Nature}\ }\textbf {\bibinfo {volume} {462}},\ \bibinfo {pages}
  {74} (\bibinfo {year} {2009})}\BibitemShut {NoStop}%
\bibitem [{\citenamefont {Sherson}\ \emph {et~al.}(2010)\citenamefont
  {Sherson}, \citenamefont {Weitenberg}, \citenamefont {Endres}, \citenamefont
  {Cheneau}, \citenamefont {Bloch},\ and\ \citenamefont {Kuhr}}]{Sherson2010}%
  \BibitemOpen
  \bibfield  {author} {\bibinfo {author} {\bibfnamefont {J.~F.}\ \bibnamefont
  {Sherson}}, \bibinfo {author} {\bibfnamefont {C.}~\bibnamefont {Weitenberg}},
  \bibinfo {author} {\bibfnamefont {M.}~\bibnamefont {Endres}}, \bibinfo
  {author} {\bibfnamefont {M.}~\bibnamefont {Cheneau}}, \bibinfo {author}
  {\bibfnamefont {I.}~\bibnamefont {Bloch}},\ and\ \bibinfo {author}
  {\bibfnamefont {S.}~\bibnamefont {Kuhr}},\ }\bibfield  {title} {\bibinfo
  {title} {Single-atom-resolved fluorescence imaging of an atomic mott
  insulator},\ }\href {https://doi.org/10.1038/nature09378} {\bibfield
  {journal} {\bibinfo  {journal} {Nature}\ }\textbf {\bibinfo {volume} {467}},\
  \bibinfo {pages} {68} (\bibinfo {year} {2010})}\BibitemShut {NoStop}%
\bibitem [{\citenamefont {An}\ \emph {et~al.}(2021)\citenamefont {An},
  \citenamefont {Padavi\ifmmode~\acute{c}\else \'{c}\fi{}}, \citenamefont
  {Meier}, \citenamefont {Hegde}, \citenamefont {Ganeshan}, \citenamefont
  {Pixley}, \citenamefont {Vishveshwara},\ and\ \citenamefont
  {Gadway}}]{Fangzhao_2020}%
  \BibitemOpen
  \bibfield  {author} {\bibinfo {author} {\bibfnamefont {F.~A.}\ \bibnamefont
  {An}}, \bibinfo {author} {\bibfnamefont {K.}~\bibnamefont
  {Padavi\ifmmode~\acute{c}\else \'{c}\fi{}}}, \bibinfo {author} {\bibfnamefont
  {E.~J.}\ \bibnamefont {Meier}}, \bibinfo {author} {\bibfnamefont
  {S.}~\bibnamefont {Hegde}}, \bibinfo {author} {\bibfnamefont
  {S.}~\bibnamefont {Ganeshan}}, \bibinfo {author} {\bibfnamefont {J.~H.}\
  \bibnamefont {Pixley}}, \bibinfo {author} {\bibfnamefont {S.}~\bibnamefont
  {Vishveshwara}},\ and\ \bibinfo {author} {\bibfnamefont {B.}~\bibnamefont
  {Gadway}},\ }\bibfield  {title} {\bibinfo {title} {Interactions and mobility
  edges: Observing the generalized aubry-andr\'e model},\ }\href
  {https://doi.org/10.1103/PhysRevLett.126.040603} {\bibfield  {journal}
  {\bibinfo  {journal} {Phys. Rev. Lett.}\ }\textbf {\bibinfo {volume} {126}},\
  \bibinfo {pages} {040603} (\bibinfo {year} {2021})}\BibitemShut {NoStop}%
\bibitem [{\citenamefont {L\"uschen}\ \emph {et~al.}(2017)\citenamefont
  {L\"uschen}, \citenamefont {Bordia}, \citenamefont {Scherg}, \citenamefont
  {Alet}, \citenamefont {Altman}, \citenamefont {Schneider},\ and\
  \citenamefont {Bloch}}]{Slow_Altman_2017}%
  \BibitemOpen
  \bibfield  {author} {\bibinfo {author} {\bibfnamefont {H.~P.}\ \bibnamefont
  {L\"uschen}}, \bibinfo {author} {\bibfnamefont {P.}~\bibnamefont {Bordia}},
  \bibinfo {author} {\bibfnamefont {S.}~\bibnamefont {Scherg}}, \bibinfo
  {author} {\bibfnamefont {F.}~\bibnamefont {Alet}}, \bibinfo {author}
  {\bibfnamefont {E.}~\bibnamefont {Altman}}, \bibinfo {author} {\bibfnamefont
  {U.}~\bibnamefont {Schneider}},\ and\ \bibinfo {author} {\bibfnamefont
  {I.}~\bibnamefont {Bloch}},\ }\bibfield  {title} {\bibinfo {title}
  {Observation of slow dynamics near the many-body localization transition in
  one-dimensional quasiperiodic systems},\ }\href
  {https://doi.org/10.1103/PhysRevLett.119.260401} {\bibfield  {journal}
  {\bibinfo  {journal} {Phys. Rev. Lett.}\ }\textbf {\bibinfo {volume} {119}},\
  \bibinfo {pages} {260401} (\bibinfo {year} {2017})}\BibitemShut {NoStop}%
\bibitem [{\citenamefont {Bordia}\ \emph {et~al.}(2017)\citenamefont {Bordia},
  \citenamefont {L\"uschen}, \citenamefont {Scherg}, \citenamefont
  {Gopalakrishnan}, \citenamefont {Knap}, \citenamefont {Schneider},\ and\
  \citenamefont {Bloch}}]{Bordia_2017}%
  \BibitemOpen
  \bibfield  {author} {\bibinfo {author} {\bibfnamefont {P.}~\bibnamefont
  {Bordia}}, \bibinfo {author} {\bibfnamefont {H.}~\bibnamefont {L\"uschen}},
  \bibinfo {author} {\bibfnamefont {S.}~\bibnamefont {Scherg}}, \bibinfo
  {author} {\bibfnamefont {S.}~\bibnamefont {Gopalakrishnan}}, \bibinfo
  {author} {\bibfnamefont {M.}~\bibnamefont {Knap}}, \bibinfo {author}
  {\bibfnamefont {U.}~\bibnamefont {Schneider}},\ and\ \bibinfo {author}
  {\bibfnamefont {I.}~\bibnamefont {Bloch}},\ }\bibfield  {title} {\bibinfo
  {title} {Probing slow relaxation and many-body localization in
  two-dimensional quasiperiodic systems},\ }\href
  {https://doi.org/10.1103/PhysRevX.7.041047} {\bibfield  {journal} {\bibinfo
  {journal} {Phys. Rev. X}\ }\textbf {\bibinfo {volume} {7}},\ \bibinfo {pages}
  {041047} (\bibinfo {year} {2017})}\BibitemShut {NoStop}%
\bibitem [{\citenamefont {Bordia}\ \emph {et~al.}(2016)\citenamefont {Bordia},
  \citenamefont {L\"uschen}, \citenamefont {Hodgman}, \citenamefont
  {Schreiber}, \citenamefont {Bloch},\ and\ \citenamefont
  {Schneider}}]{Bordia_2016}%
  \BibitemOpen
  \bibfield  {author} {\bibinfo {author} {\bibfnamefont {P.}~\bibnamefont
  {Bordia}}, \bibinfo {author} {\bibfnamefont {H.~P.}\ \bibnamefont
  {L\"uschen}}, \bibinfo {author} {\bibfnamefont {S.~S.}\ \bibnamefont
  {Hodgman}}, \bibinfo {author} {\bibfnamefont {M.}~\bibnamefont {Schreiber}},
  \bibinfo {author} {\bibfnamefont {I.}~\bibnamefont {Bloch}},\ and\ \bibinfo
  {author} {\bibfnamefont {U.}~\bibnamefont {Schneider}},\ }\bibfield  {title}
  {\bibinfo {title} {Coupling identical one-dimensional many-body localized
  systems},\ }\href {https://doi.org/10.1103/PhysRevLett.116.140401} {\bibfield
   {journal} {\bibinfo  {journal} {Phys. Rev. Lett.}\ }\textbf {\bibinfo
  {volume} {116}},\ \bibinfo {pages} {140401} (\bibinfo {year}
  {2016})}\BibitemShut {NoStop}%
\bibitem [{\citenamefont {Kohlert}\ \emph {et~al.}(2019)\citenamefont
  {Kohlert}, \citenamefont {Scherg}, \citenamefont {Li}, \citenamefont
  {L\"uschen}, \citenamefont {Das~Sarma}, \citenamefont {Bloch},\ and\
  \citenamefont {Aidelsburger}}]{Kohlert_2019}%
  \BibitemOpen
  \bibfield  {author} {\bibinfo {author} {\bibfnamefont {T.}~\bibnamefont
  {Kohlert}}, \bibinfo {author} {\bibfnamefont {S.}~\bibnamefont {Scherg}},
  \bibinfo {author} {\bibfnamefont {X.}~\bibnamefont {Li}}, \bibinfo {author}
  {\bibfnamefont {H.~P.}\ \bibnamefont {L\"uschen}}, \bibinfo {author}
  {\bibfnamefont {S.}~\bibnamefont {Das~Sarma}}, \bibinfo {author}
  {\bibfnamefont {I.}~\bibnamefont {Bloch}},\ and\ \bibinfo {author}
  {\bibfnamefont {M.}~\bibnamefont {Aidelsburger}},\ }\bibfield  {title}
  {\bibinfo {title} {Observation of many-body localization in a one-dimensional
  system with a single-particle mobility edge},\ }\href
  {https://doi.org/10.1103/PhysRevLett.122.170403} {\bibfield  {journal}
  {\bibinfo  {journal} {Phys. Rev. Lett.}\ }\textbf {\bibinfo {volume} {122}},\
  \bibinfo {pages} {170403} (\bibinfo {year} {2019})}\BibitemShut {NoStop}%
\bibitem [{\citenamefont {Anderson}(1958)}]{Anderson1958Loc}%
  \BibitemOpen
  \bibfield  {author} {\bibinfo {author} {\bibfnamefont {P.~W.}\ \bibnamefont
  {Anderson}},\ }\bibfield  {title} {\bibinfo {title} {Absence of diffusion in
  certain random lattices},\ }\href {https://doi.org/10.1103/PhysRev.109.1492}
  {\bibfield  {journal} {\bibinfo  {journal} {Phys. Rev.}\ }\textbf {\bibinfo
  {volume} {109}},\ \bibinfo {pages} {1492} (\bibinfo {year}
  {1958})}\BibitemShut {NoStop}%
\bibitem [{\citenamefont {Evers}\ and\ \citenamefont
  {Mirlin}(2008)}]{Evers2008Review}%
  \BibitemOpen
  \bibfield  {author} {\bibinfo {author} {\bibfnamefont {F.}~\bibnamefont
  {Evers}}\ and\ \bibinfo {author} {\bibfnamefont {A.~D.}\ \bibnamefont
  {Mirlin}},\ }\bibfield  {title} {\bibinfo {title} {Anderson transitions},\
  }\href {https://doi.org/10.1103/RevModPhys.80.1355} {\bibfield  {journal}
  {\bibinfo  {journal} {Rev. Mod. Phys.}\ }\textbf {\bibinfo {volume} {80}},\
  \bibinfo {pages} {1355} (\bibinfo {year} {2008})}\BibitemShut {NoStop}%
\bibitem [{\citenamefont {Mott}\ and\ \citenamefont
  {Twose}(1961)}]{Mott1961theory}%
  \BibitemOpen
  \bibfield  {author} {\bibinfo {author} {\bibfnamefont {N.~F.}\ \bibnamefont
  {Mott}}\ and\ \bibinfo {author} {\bibfnamefont {W.}~\bibnamefont {Twose}},\
  }\bibfield  {title} {\bibinfo {title} {{The theory of impurity conduction}},\
  }\href@noop {} {\bibfield  {journal} {\bibinfo  {journal} {Advances in
  Physics}\ }\textbf {\bibinfo {volume} {10}},\ \bibinfo {pages} {107}
  (\bibinfo {year} {1961})}\BibitemShut {NoStop}%
\bibitem [{\citenamefont {Luitz}\ \emph {et~al.}(2015)\citenamefont {Luitz},
  \citenamefont {Laflorencie},\ and\ \citenamefont {Alet}}]{Luitz15Edge}%
  \BibitemOpen
  \bibfield  {author} {\bibinfo {author} {\bibfnamefont {D.~J.}\ \bibnamefont
  {Luitz}}, \bibinfo {author} {\bibfnamefont {N.}~\bibnamefont {Laflorencie}},\
  and\ \bibinfo {author} {\bibfnamefont {F.}~\bibnamefont {Alet}},\ }\bibfield
  {title} {\bibinfo {title} {Many-body localization edge in the random-field
  heisenberg chain},\ }\href {https://doi.org/10.1103/PhysRevB.91.081103}
  {\bibfield  {journal} {\bibinfo  {journal} {Phys. Rev. B}\ }\textbf {\bibinfo
  {volume} {91}},\ \bibinfo {pages} {081103} (\bibinfo {year}
  {2015})}\BibitemShut {NoStop}%
\bibitem [{\citenamefont {De~Tomasi}\ \emph
  {et~al.}(2017{\natexlab{b}})\citenamefont {De~Tomasi}, \citenamefont {Bera},
  \citenamefont {Bardarson},\ and\ \citenamefont {Pollmann}}]{Tomasi17Mutual}%
  \BibitemOpen
  \bibfield  {author} {\bibinfo {author} {\bibfnamefont {G.}~\bibnamefont
  {De~Tomasi}}, \bibinfo {author} {\bibfnamefont {S.}~\bibnamefont {Bera}},
  \bibinfo {author} {\bibfnamefont {J.~H.}\ \bibnamefont {Bardarson}},\ and\
  \bibinfo {author} {\bibfnamefont {F.}~\bibnamefont {Pollmann}},\ }\bibfield
  {title} {\bibinfo {title} {Quantum mutual information as a probe for
  many-body localization},\ }\href
  {https://doi.org/10.1103/PhysRevLett.118.016804} {\bibfield  {journal}
  {\bibinfo  {journal} {Phys. Rev. Lett.}\ }\textbf {\bibinfo {volume} {118}},\
  \bibinfo {pages} {016804} (\bibinfo {year} {2017}{\natexlab{b}})}\BibitemShut
  {NoStop}%
\bibitem [{\citenamefont {Bera}\ \emph {et~al.}(2015)\citenamefont {Bera},
  \citenamefont {Schomerus}, \citenamefont {Heidrich-Meisner},\ and\
  \citenamefont {Bardarson}}]{Bera15OnePart}%
  \BibitemOpen
  \bibfield  {author} {\bibinfo {author} {\bibfnamefont {S.}~\bibnamefont
  {Bera}}, \bibinfo {author} {\bibfnamefont {H.}~\bibnamefont {Schomerus}},
  \bibinfo {author} {\bibfnamefont {F.}~\bibnamefont {Heidrich-Meisner}},\ and\
  \bibinfo {author} {\bibfnamefont {J.~H.}\ \bibnamefont {Bardarson}},\
  }\bibfield  {title} {\bibinfo {title} {Many-body localization characterized
  from a one-particle perspective},\ }\href
  {https://doi.org/10.1103/PhysRevLett.115.046603} {\bibfield  {journal}
  {\bibinfo  {journal} {Phys. Rev. Lett.}\ }\textbf {\bibinfo {volume} {115}},\
  \bibinfo {pages} {046603} (\bibinfo {year} {2015})}\BibitemShut {NoStop}%
\bibitem [{\citenamefont {Serbyn}\ \emph {et~al.}(2015)\citenamefont {Serbyn},
  \citenamefont {Papi{\'c}},\ and\ \citenamefont
  {Abanin}}]{Serbyn2015criterion}%
  \BibitemOpen
  \bibfield  {author} {\bibinfo {author} {\bibfnamefont {M.}~\bibnamefont
  {Serbyn}}, \bibinfo {author} {\bibfnamefont {Z.}~\bibnamefont {Papi{\'c}}},\
  and\ \bibinfo {author} {\bibfnamefont {D.~A.}\ \bibnamefont {Abanin}},\
  }\bibfield  {title} {\bibinfo {title} {{Criterion for Many-Body
  Localization-Delocalization Phase Transition}},\ }\href@noop {} {\bibfield
  {journal} {\bibinfo  {journal} {Phys. Rev. X}\ }\textbf {\bibinfo {volume}
  {5}},\ \bibinfo {pages} {041047} (\bibinfo {year} {2015})}\BibitemShut
  {NoStop}%
\bibitem [{\citenamefont {Pal}\ and\ \citenamefont {Huse}(2010)}]{Pal_2010}%
  \BibitemOpen
  \bibfield  {author} {\bibinfo {author} {\bibfnamefont {A.}~\bibnamefont
  {Pal}}\ and\ \bibinfo {author} {\bibfnamefont {D.~A.}\ \bibnamefont {Huse}},\
  }\bibfield  {title} {\bibinfo {title} {Many-body localization phase
  transition},\ }\href {https://doi.org/10.1103/PhysRevB.82.174411} {\bibfield
  {journal} {\bibinfo  {journal} {Phys. Rev. B}\ }\textbf {\bibinfo {volume}
  {82}},\ \bibinfo {pages} {174411} (\bibinfo {year} {2010})}\BibitemShut
  {NoStop}%
\bibitem [{\citenamefont {Oganesyan}\ and\ \citenamefont
  {Huse}(2007)}]{Oga_2007}%
  \BibitemOpen
  \bibfield  {author} {\bibinfo {author} {\bibfnamefont {V.}~\bibnamefont
  {Oganesyan}}\ and\ \bibinfo {author} {\bibfnamefont {D.~A.}\ \bibnamefont
  {Huse}},\ }\bibfield  {title} {\bibinfo {title} {Localization of interacting
  fermions at high temperature},\ }\href
  {https://doi.org/10.1103/PhysRevB.75.155111} {\bibfield  {journal} {\bibinfo
  {journal} {Phys. Rev. B}\ }\textbf {\bibinfo {volume} {75}},\ \bibinfo
  {pages} {155111} (\bibinfo {year} {2007})}\BibitemShut {NoStop}%
\bibitem [{\citenamefont {Aubry}\ and\ \citenamefont
  {André}(1980)}]{Aubrey_1980}%
  \BibitemOpen
  \bibfield  {author} {\bibinfo {author} {\bibfnamefont {S.}~\bibnamefont
  {Aubry}}\ and\ \bibinfo {author} {\bibfnamefont {G.}~\bibnamefont {André}},\
  }\bibfield  {title} {\bibinfo {title} {Analyticity breaking and anderson
  localization in incommensurate lattices},\ }\href@noop {} {\bibfield
  {journal} {\bibinfo  {journal} {Ann. Israel Phys. Soc}\ }\textbf {\bibinfo
  {volume} {3}},\ \bibinfo {pages} {18} (\bibinfo {year} {1980})}\BibitemShut
  {NoStop}%
\bibitem [{\citenamefont {Iyer}\ \emph {et~al.}(2013)\citenamefont {Iyer},
  \citenamefont {Oganesyan}, \citenamefont {Refael},\ and\ \citenamefont
  {Huse}}]{Shankar_2012}%
  \BibitemOpen
  \bibfield  {author} {\bibinfo {author} {\bibfnamefont {S.}~\bibnamefont
  {Iyer}}, \bibinfo {author} {\bibfnamefont {V.}~\bibnamefont {Oganesyan}},
  \bibinfo {author} {\bibfnamefont {G.}~\bibnamefont {Refael}},\ and\ \bibinfo
  {author} {\bibfnamefont {D.~A.}\ \bibnamefont {Huse}},\ }\bibfield  {title}
  {\bibinfo {title} {Many-body localization in a quasiperiodic system},\ }\href
  {https://doi.org/10.1103/PhysRevB.87.134202} {\bibfield  {journal} {\bibinfo
  {journal} {Phys. Rev. B}\ }\textbf {\bibinfo {volume} {87}},\ \bibinfo
  {pages} {134202} (\bibinfo {year} {2013})}\BibitemShut {NoStop}%
\bibitem [{\citenamefont {Rawat}\ and\ \citenamefont
  {Wang}(2017)}]{rawat2017deep}%
  \BibitemOpen
  \bibfield  {author} {\bibinfo {author} {\bibfnamefont {W.}~\bibnamefont
  {Rawat}}\ and\ \bibinfo {author} {\bibfnamefont {Z.}~\bibnamefont {Wang}},\
  }\bibfield  {title} {\bibinfo {title} {Deep convolutional neural networks for
  image classification: A comprehensive review},\ }\href@noop {} {\bibfield
  {journal} {\bibinfo  {journal} {Neural computation}\ }\textbf {\bibinfo
  {volume} {29}},\ \bibinfo {pages} {2352} (\bibinfo {year}
  {2017})}\BibitemShut {NoStop}%
\bibitem [{\citenamefont {Szegedy}\ \emph {et~al.}(2014)\citenamefont
  {Szegedy}, \citenamefont {Liu}, \citenamefont {Jia}, \citenamefont
  {Sermanet}, \citenamefont {Reed}, \citenamefont {Anguelov}, \citenamefont
  {Erhan}, \citenamefont {Vanhoucke},\ and\ \citenamefont
  {Rabinovich}}]{Szegedy2014Inception}%
  \BibitemOpen
  \bibfield  {author} {\bibinfo {author} {\bibfnamefont {C.}~\bibnamefont
  {Szegedy}}, \bibinfo {author} {\bibfnamefont {W.}~\bibnamefont {Liu}},
  \bibinfo {author} {\bibfnamefont {Y.}~\bibnamefont {Jia}}, \bibinfo {author}
  {\bibfnamefont {P.}~\bibnamefont {Sermanet}}, \bibinfo {author}
  {\bibfnamefont {S.}~\bibnamefont {Reed}}, \bibinfo {author} {\bibfnamefont
  {D.}~\bibnamefont {Anguelov}}, \bibinfo {author} {\bibfnamefont
  {D.}~\bibnamefont {Erhan}}, \bibinfo {author} {\bibfnamefont
  {V.}~\bibnamefont {Vanhoucke}},\ and\ \bibinfo {author} {\bibfnamefont
  {A.}~\bibnamefont {Rabinovich}},\ }\href@noop {} {\bibinfo {title} {Going
  deeper with convolutions}} (\bibinfo {year} {2014}),\ \Eprint
  {https://arxiv.org/abs/1409.4842} {arXiv:1409.4842 [cs.CV]} \BibitemShut
  {NoStop}%
\bibitem [{\citenamefont {Goodfellow}\ \emph {et~al.}(2016)\citenamefont
  {Goodfellow}, \citenamefont {Bengio},\ and\ \citenamefont
  {Courville}}]{Goodfellow2016DLBook}%
  \BibitemOpen
  \bibfield  {author} {\bibinfo {author} {\bibfnamefont {I.}~\bibnamefont
  {Goodfellow}}, \bibinfo {author} {\bibfnamefont {Y.}~\bibnamefont {Bengio}},\
  and\ \bibinfo {author} {\bibfnamefont {A.}~\bibnamefont {Courville}},\
  }\href@noop {} {\emph {\bibinfo {title} {Deep Learning}}}\ (\bibinfo
  {publisher} {MIT Press},\ \bibinfo {year} {2016})\ \bibinfo {note}
  {\url{http://www.deeplearningbook.org}}\BibitemShut {NoStop}%
\bibitem [{\citenamefont {Serbyn}\ \emph
  {et~al.}(2014{\natexlab{c}})\citenamefont {Serbyn}, \citenamefont
  {Papi\ifmmode~\acute{c}\else \'{c}\fi{}},\ and\ \citenamefont
  {Abanin}}]{Serbyn_2014_local}%
  \BibitemOpen
  \bibfield  {author} {\bibinfo {author} {\bibfnamefont {M.}~\bibnamefont
  {Serbyn}}, \bibinfo {author} {\bibfnamefont {Z.}~\bibnamefont
  {Papi\ifmmode~\acute{c}\else \'{c}\fi{}}},\ and\ \bibinfo {author}
  {\bibfnamefont {D.~A.}\ \bibnamefont {Abanin}},\ }\bibfield  {title}
  {\bibinfo {title} {Quantum quenches in the many-body localized phase},\
  }\href {https://doi.org/10.1103/PhysRevB.90.174302} {\bibfield  {journal}
  {\bibinfo  {journal} {Phys. Rev. B}\ }\textbf {\bibinfo {volume} {90}},\
  \bibinfo {pages} {174302} (\bibinfo {year} {2014}{\natexlab{c}})}\BibitemShut
  {NoStop}%
\bibitem [{Note1()}]{Note1}%
  \BibitemOpen
  \bibinfo {note} {The target times are log-linear distributed.}\BibitemShut
  {Stop}%
\bibitem [{\citenamefont {{Scherg}}\ \emph {et~al.}(2020)\citenamefont
  {{Scherg}}, \citenamefont {{Kohlert}}, \citenamefont {{Sala}}, \citenamefont
  {{Pollmann}}, \citenamefont {{Bharath H.}}, \citenamefont {{Bloch}},\ and\
  \citenamefont {{Aidelsburger}}}]{Aidelsburger_2020}%
  \BibitemOpen
  \bibfield  {author} {\bibinfo {author} {\bibfnamefont {S.}~\bibnamefont
  {{Scherg}}}, \bibinfo {author} {\bibfnamefont {T.}~\bibnamefont {{Kohlert}}},
  \bibinfo {author} {\bibfnamefont {P.}~\bibnamefont {{Sala}}}, \bibinfo
  {author} {\bibfnamefont {F.}~\bibnamefont {{Pollmann}}}, \bibinfo {author}
  {\bibfnamefont {M.}~\bibnamefont {{Bharath H.}}}, \bibinfo {author}
  {\bibfnamefont {I.}~\bibnamefont {{Bloch}}},\ and\ \bibinfo {author}
  {\bibfnamefont {M.}~\bibnamefont {{Aidelsburger}}},\ }\bibfield  {title}
  {\bibinfo {title} {{Observing non-ergodicity due to kinetic constraints in
  tilted Fermi-Hubbard chains}},\ }\href@noop {} {\bibfield  {journal}
  {\bibinfo  {journal} {arXiv e-prints}\ ,\ \bibinfo {eid} {arXiv:2010.12965}}
  (\bibinfo {year} {2020})},\ \Eprint {https://arxiv.org/abs/2010.12965}
  {arXiv:2010.12965 [cond-mat.quant-gas]} \BibitemShut {NoStop}%
\bibitem [{\citenamefont {\v{Z}nidari\v{c}}\ \emph {et~al.}(2008)\citenamefont
  {\v{Z}nidari\v{c}}, \citenamefont {Prosen},\ and\ \citenamefont
  {Prelov\v{s}ek}}]{Vznidarivc2008many}%
  \BibitemOpen
  \bibfield  {author} {\bibinfo {author} {\bibfnamefont {M.}~\bibnamefont
  {\v{Z}nidari\v{c}}}, \bibinfo {author} {\bibfnamefont {T.}~\bibnamefont
  {Prosen}},\ and\ \bibinfo {author} {\bibfnamefont {P.}~\bibnamefont
  {Prelov\v{s}ek}},\ }\bibfield  {title} {\bibinfo {title} {{Many-body
  localization in the {Heisenberg XXZ} magnet in a random field}},\ }\href
  {https://journals.aps.org/prb/abstract/10.1103/PhysRevB.77.064426} {\bibfield
   {journal} {\bibinfo  {journal} {Phys. Rev. B}\ }\textbf {\bibinfo {volume}
  {77}},\ \bibinfo {pages} {064426} (\bibinfo {year} {2008})}\BibitemShut
  {NoStop}%
\bibitem [{Note2()}]{Note2}%
  \BibitemOpen
  \bibinfo {note} {This argument can work only if the localization length of of
  the system is larger than the subsystem.}\BibitemShut {Stop}%
\bibitem [{Note3()}]{Note3}%
  \BibitemOpen
  \bibinfo {note} {Fraction of correctly classified MBL $= \protect \frac { \#
  \protect \textit {MBL classified} }{\# \protect \textit {
  MBL}}$.}\BibitemShut {Stop}%
\bibitem [{Note4()}]{Note4}%
  \BibitemOpen
  \bibinfo {note} {Additionally, we exploit the fact that we can produce more
  cut-outs from larger systems, simply by using subsystems ($n_{0},\protect
  \cdots , n_{\ell -1}$) for the first cut-out, then ($n_{1},\protect \cdots ,
  n_{\ell }$) for the second and so on until we reach the end of the chain
  ($n_{L-\ell },\protect \cdots , n_{L-1}$). In testing, all cut-outs of one
  snapshot block are classified and labeled as one of the two phases. After
  going through all cut-outs of one snapshot block of length $L$, one assigns
  the category to the whole snapshot block which was ascribed to a majority of
  its cut-outs. We call this procedure a voting mechanism, since each
  classification of cut-outs gets one vote, and the majority vote decides which
  label is assigned to the whole system.}\BibitemShut {Stop}%
\bibitem [{\citenamefont {Mordvintsev}\ \emph {et~al.}(2015)\citenamefont
  {Mordvintsev}, \citenamefont {Olah},\ and\ \citenamefont
  {Tyka}}]{Mordvintsev2015Dreaming}%
  \BibitemOpen
  \bibfield  {author} {\bibinfo {author} {\bibfnamefont {A.}~\bibnamefont
  {Mordvintsev}}, \bibinfo {author} {\bibfnamefont {C.}~\bibnamefont {Olah}},\
  and\ \bibinfo {author} {\bibfnamefont {M.}~\bibnamefont {Tyka}},\ }\href@noop
  {} {\bibinfo {title} {Inceptionism: Going deeper into neural networks}},\
  \bibinfo {howpublished} {Google AI Blog} (\bibinfo {year} {2015})\BibitemShut
  {NoStop}%
\bibitem [{\citenamefont {Luitz}\ and\ \citenamefont
  {Lev}(2020)}]{Bar_Lev_2020}%
  \BibitemOpen
  \bibfield  {author} {\bibinfo {author} {\bibfnamefont {D.~J.}\ \bibnamefont
  {Luitz}}\ and\ \bibinfo {author} {\bibfnamefont {Y.~B.}\ \bibnamefont
  {Lev}},\ }\bibfield  {title} {\bibinfo {title} {Absence of slow particle
  transport in the many-body localized phase},\ }\href
  {https://doi.org/10.1103/PhysRevB.102.100202} {\bibfield  {journal} {\bibinfo
   {journal} {Phys. Rev. B}\ }\textbf {\bibinfo {volume} {102}},\ \bibinfo
  {pages} {100202} (\bibinfo {year} {2020})}\BibitemShut {NoStop}%
\bibitem [{\citenamefont {Kiefer-Emmanouilidis}\ \emph
  {et~al.}(2021)\citenamefont {Kiefer-Emmanouilidis}, \citenamefont {Unanyan},
  \citenamefont {Fleischhauer},\ and\ \citenamefont {Sirker}}]{Jesko_2021}%
  \BibitemOpen
  \bibfield  {author} {\bibinfo {author} {\bibfnamefont {M.}~\bibnamefont
  {Kiefer-Emmanouilidis}}, \bibinfo {author} {\bibfnamefont {R.}~\bibnamefont
  {Unanyan}}, \bibinfo {author} {\bibfnamefont {M.}~\bibnamefont
  {Fleischhauer}},\ and\ \bibinfo {author} {\bibfnamefont {J.}~\bibnamefont
  {Sirker}},\ }\bibfield  {title} {\bibinfo {title} {Slow delocalization of
  particles in many-body localized phases},\ }\href
  {https://doi.org/10.1103/PhysRevB.103.024203} {\bibfield  {journal} {\bibinfo
   {journal} {Phys. Rev. B}\ }\textbf {\bibinfo {volume} {103}},\ \bibinfo
  {pages} {024203} (\bibinfo {year} {2021})}\BibitemShut {NoStop}%
\bibitem [{\citenamefont {Lee}\ \emph {et~al.}(2017)\citenamefont {Lee},
  \citenamefont {Look}, \citenamefont {Lim},\ and\ \citenamefont
  {Sheng}}]{Lim_2017}%
  \BibitemOpen
  \bibfield  {author} {\bibinfo {author} {\bibfnamefont {M.}~\bibnamefont
  {Lee}}, \bibinfo {author} {\bibfnamefont {T.~R.}\ \bibnamefont {Look}},
  \bibinfo {author} {\bibfnamefont {S.~P.}\ \bibnamefont {Lim}},\ and\ \bibinfo
  {author} {\bibfnamefont {D.~N.}\ \bibnamefont {Sheng}},\ }\bibfield  {title}
  {\bibinfo {title} {Many-body localization in spin chain systems with
  quasiperiodic fields},\ }\href {https://doi.org/10.1103/PhysRevB.96.075146}
  {\bibfield  {journal} {\bibinfo  {journal} {Phys. Rev. B}\ }\textbf {\bibinfo
  {volume} {96}},\ \bibinfo {pages} {075146} (\bibinfo {year}
  {2017})}\BibitemShut {NoStop}%
\bibitem [{\citenamefont {{Singh}}\ \emph {et~al.}(2021)\citenamefont
  {{Singh}}, \citenamefont {{Ware}}, \citenamefont {{Vasseur}},\ and\
  \citenamefont {{Gopalakrishnan}}}]{Sarang_2021}%
  \BibitemOpen
  \bibfield  {author} {\bibinfo {author} {\bibfnamefont {H.}~\bibnamefont
  {{Singh}}}, \bibinfo {author} {\bibfnamefont {B.}~\bibnamefont {{Ware}}},
  \bibinfo {author} {\bibfnamefont {R.}~\bibnamefont {{Vasseur}}},\ and\
  \bibinfo {author} {\bibfnamefont {S.}~\bibnamefont {{Gopalakrishnan}}},\
  }\bibfield  {title} {\bibinfo {title} {{Local integrals of motion and the
  quasiperiodic many-body localization transition}},\ }\href@noop {} {\bibfield
   {journal} {\bibinfo  {journal} {arXiv e-prints}\ ,\ \bibinfo {eid}
  {arXiv:2101.04126}} (\bibinfo {year} {2021})},\ \Eprint
  {https://arxiv.org/abs/2101.04126} {arXiv:2101.04126 [cond-mat.dis-nn]}
  \BibitemShut {NoStop}%
\bibitem [{cas(1995)}]{casati_chirikov_1995}%
  \BibitemOpen
  \href {https://doi.org/10.1017/CBO9780511599989} {\emph {\bibinfo {title}
  {Quantum Chaos: Between Order and Disorder}}}\ (\bibinfo  {publisher}
  {Cambridge University Press},\ \bibinfo {year} {1995})\BibitemShut {NoStop}%
\bibitem [{\citenamefont {Ros}\ \emph {et~al.}(2015)\citenamefont {Ros},
  \citenamefont {M{\"u}ller},\ and\ \citenamefont
  {Scardicchio}}]{Ros2015Integrals}%
  \BibitemOpen
  \bibfield  {author} {\bibinfo {author} {\bibfnamefont {V.}~\bibnamefont
  {Ros}}, \bibinfo {author} {\bibfnamefont {M.}~\bibnamefont {M{\"u}ller}},\
  and\ \bibinfo {author} {\bibfnamefont {A.}~\bibnamefont {Scardicchio}},\
  }\bibfield  {title} {\bibinfo {title} {{Integrals of motion in the many-body
  localized phase}},\ }\href@noop {} {\bibfield  {journal} {\bibinfo  {journal}
  {Nuclear Physics B}\ }\textbf {\bibinfo {volume} {891}},\ \bibinfo {pages}
  {420} (\bibinfo {year} {2015})},\ \bibinfo {note} {[Corrigendum.
  \emph{Nuclear Physics B}, 900:446 - 448, 2015]}\BibitemShut {NoStop}%
\bibitem [{\citenamefont {Kingma}\ and\ \citenamefont
  {Ba}(2014)}]{Kingma2014Adam}%
  \BibitemOpen
  \bibfield  {author} {\bibinfo {author} {\bibfnamefont {D.~P.}\ \bibnamefont
  {Kingma}}\ and\ \bibinfo {author} {\bibfnamefont {J.}~\bibnamefont {Ba}},\
  }\href@noop {} {\bibinfo {title} {Adam: A method for stochastic
  optimization}} (\bibinfo {year} {2014}),\ \Eprint
  {https://arxiv.org/abs/1412.6980} {arXiv:1412.6980 [cs.LG]} \BibitemShut
  {NoStop}%
\bibitem [{\citenamefont {Paszke}\ \emph {et~al.}(2019)\citenamefont {Paszke},
  \citenamefont {Gross}, \citenamefont {Massa}, \citenamefont {Lerer},
  \citenamefont {Bradbury}, \citenamefont {Chanan}, \citenamefont {Killeen},
  \citenamefont {Lin}, \citenamefont {Gimelshein}, \citenamefont {Antiga},
  \citenamefont {Desmaison}, \citenamefont {Kopf}, \citenamefont {Yang},
  \citenamefont {DeVito}, \citenamefont {Raison}, \citenamefont {Tejani},
  \citenamefont {Chilamkurthy}, \citenamefont {Steiner}, \citenamefont {Fang},
  \citenamefont {Bai},\ and\ \citenamefont {Chintala}}]{Paszke2019Pytorch}%
  \BibitemOpen
  \bibfield  {author} {\bibinfo {author} {\bibfnamefont {A.}~\bibnamefont
  {Paszke}}, \bibinfo {author} {\bibfnamefont {S.}~\bibnamefont {Gross}},
  \bibinfo {author} {\bibfnamefont {F.}~\bibnamefont {Massa}}, \bibinfo
  {author} {\bibfnamefont {A.}~\bibnamefont {Lerer}}, \bibinfo {author}
  {\bibfnamefont {J.}~\bibnamefont {Bradbury}}, \bibinfo {author}
  {\bibfnamefont {G.}~\bibnamefont {Chanan}}, \bibinfo {author} {\bibfnamefont
  {T.}~\bibnamefont {Killeen}}, \bibinfo {author} {\bibfnamefont
  {Z.}~\bibnamefont {Lin}}, \bibinfo {author} {\bibfnamefont {N.}~\bibnamefont
  {Gimelshein}}, \bibinfo {author} {\bibfnamefont {L.}~\bibnamefont {Antiga}},
  \bibinfo {author} {\bibfnamefont {A.}~\bibnamefont {Desmaison}}, \bibinfo
  {author} {\bibfnamefont {A.}~\bibnamefont {Kopf}}, \bibinfo {author}
  {\bibfnamefont {E.}~\bibnamefont {Yang}}, \bibinfo {author} {\bibfnamefont
  {Z.}~\bibnamefont {DeVito}}, \bibinfo {author} {\bibfnamefont
  {M.}~\bibnamefont {Raison}}, \bibinfo {author} {\bibfnamefont
  {A.}~\bibnamefont {Tejani}}, \bibinfo {author} {\bibfnamefont
  {S.}~\bibnamefont {Chilamkurthy}}, \bibinfo {author} {\bibfnamefont
  {B.}~\bibnamefont {Steiner}}, \bibinfo {author} {\bibfnamefont
  {L.}~\bibnamefont {Fang}}, \bibinfo {author} {\bibfnamefont {J.}~\bibnamefont
  {Bai}},\ and\ \bibinfo {author} {\bibfnamefont {S.}~\bibnamefont
  {Chintala}},\ }\bibfield  {title} {\bibinfo {title} {Pytorch: An imperative
  style, high-performance deep learning library},\ }in\ \href
  {http://papers.nips.cc/paper/9015-pytorch-an-imperative-style-high-performance-deep-learning-library.pdf}
  {\emph {\bibinfo {booktitle} {Advances in Neural Information Processing
  Systems 32}}},\ \bibinfo {editor} {edited by\ \bibinfo {editor}
  {\bibfnamefont {H.}~\bibnamefont {Wallach}}, \bibinfo {editor} {\bibfnamefont
  {H.}~\bibnamefont {Larochelle}}, \bibinfo {editor} {\bibfnamefont
  {A.}~\bibnamefont {Beygelzimer}}, \bibinfo {editor} {\bibfnamefont
  {F.}~\bibnamefont {d\textquotesingle Alch\'{e}-Buc}}, \bibinfo {editor}
  {\bibfnamefont {E.}~\bibnamefont {Fox}},\ and\ \bibinfo {editor}
  {\bibfnamefont {R.}~\bibnamefont {Garnett}}}\ (\bibinfo  {publisher} {Curran
  Associates, Inc.},\ \bibinfo {year} {2019})\ pp.\ \bibinfo {pages}
  {8026--8037}\BibitemShut {NoStop}%
\end{thebibliography}%

\end{document}